\title{CDS RATE CONSTRUCTION METHODS \\by
 MACHINE LEARNING TECHNIQUES}
\author{Raymond Brummelhuis\thanks{University of Reims, Department of Mathematics and Computer Science, Reims, France %and Birkbeck, University of London
                  }
        \and
        Zhongmin Luo\thanks{
                      Birkbeck, University of London
                      and Standard Chartered Bank; the views expressed in the paper are the author's own and do not necessarily reflect those of the author's affiliated institutions.
                  }
        }
\documentclass{article}
\usepackage{graphicx}
\usepackage{latexsym,amsmath}
\usepackage{amsfonts}
\usepackage{amssymb}
\usepackage{amsthm}
\usepackage{pifont} %% for \ding
\usepackage{hhline}
\usepackage{dsfont}
\usepackage{epsfig}
\usepackage{rotating}
\usepackage{url}
\usepackage{caption}
\usepackage{makeidx}
\usepackage{bm}
\usepackage{hyperref}
\makeindex

\usepackage{txfonts} 
\usepackage{pxfonts}
\usepackage[titletoc,title]{appendix}
\usepackage{mathtools}
\usepackage[top=1in, bottom=1.25in, left=1.35in, right=1.20in]{geometry}
\usepackage{geometry}
\theoremstyle{definition}

\newenvironment{keywords}{}{}
\begin{document}

%\jvol{00} \jnum{00} \jyear{2014} \jmonth{October}

\title{CDS RATE CONSTRUCTION METHODS \\by
 MACHINE LEARNING TECHNIQUES}

\author{Raymond Brummelhuis${\dag}$ and Zhongmin Luo$^{\ast}$${\ddag}$\thanks{$^\ast$Corresponding author.Email: zhongmin.luo@btinternet.com}\\\affil{
$\dag$Universit\'e de Reims-Champagne-Ardenne, Laboratoire de Math\'ematiques EA 4535, Reims, France\\
$\ddag$Birkbeck, University of London and Standard Chartered Bank Plc, Group Analytics, London, UK} \received{v1.0 released April 2017} }
  
\title{CDS RATE CONSTRUCTION METHODS \\by
 MACHINE LEARNING TECHNIQUES}
\author{Raymond Brummelhuis\thanks{Universit\'e de Reims-Champagne-Ardenne, Laboratoire de Math\'ematiques EA 4535, Reims, France %and Birkbeck, University of London
                  }
        \and
        Zhongmin Luo\thanks{
                      Birkbeck, University of London
                      and Standard Chartered Bank; the views expressed in the paper are the author's own and do not necessarily reflect those of the author's affiliated institutions.
                  }
        }

%\author{A. N. AUTHOR$^{\ast}$$\dag$\thanks{$^\ast$Corresponding author.
%Email: latex.helpdesk@tandf.co.uk} and I. T. CONSULTANT${\ddag}$\\
%\affil{$\dag$Taylor \& Francis, 4 Park Square, Milton Park, Abingdon, OX14 4RN, UK\\
%$\ddag$Institut f\"{u}r Informatik, Albert-Ludwigs-Universit\"{a}t,
%D-79110 Freiburg, Germany} \received{v2.1 released October 2014} }

\maketitle

\begin{abstract}
Regulators require financial institutions to estimate counterparty default risks from liquid
CDS quotes for the valuation and risk management of OTC derivatives. However, the vast majority of counterparties do not have liquid CDS quotes and need proxy CDS rates. Existing
methods cannot account for counterparty-specific default risks; we propose to construct proxy
CDS rates by associating to illiquid counterparty liquid CDS Proxy based on Machine Learning
Techniques. After testing 156 classifiers from 8 most popular classifier families, we found that
some classifiers achieve highly satisfactory accuracy rates. Furthermore, we have rank-ordered
the performances and investigated performance variations amongst and within the 8 classifier
families. This paper is, to the best of our knowledge, the first systematic study of CDS Proxy
construction by Machine Learning techniques, and the first systematic classifier comparison
study based entirely on financial market data. Its findings both confirm and contrast existing
classifier performance literature. Given the typically highly correlated nature of financial data, we investigated
the impact of correlation on classifier performance. The techniques used in this paper should
be of interest for financial institutions seeking a CDS Proxy method, and can serve for proxy
construction for other financial variables. Some directions for future research are indicated.
\end{abstract}


\section{Introduction}\label{sectionIntroduction}
\subsection{A Shortage of Liquidity Problem}\label{probliqidity}   One important lesson learned from the 2008 financial crisis is that the valuation of Over-the-Counter (OTC) derivatives in financial institutions did not correctly take  into account the risk of default associated with counterparties, the so-called \textit{Counterparty Credit Risk} (Brigo {\it et al.} 2103). %\cite{Brigo}.   
To align the risk neutral valuation of OTC derivatives with the risks to which investors are exposed, the finance industry has come to recognize that it is critical to make adjustments to the default-free valuation of OTC derivatives using metrics such as Credit Value Adjustment (CVA), Funding Value Adjustment (FVA), Collateral Value Adjustment (ColVA) and Capital Value Adjustment (KVA) (altogether often referred to as XVA, see Gregory, 2015).

In 2010, the Basel Committee on Banking Supervision (BCBS) published Basel III (BCBS, 2010)   
which requires banks to provide Value-at-Risk-based capital reserves against CVA losses, to account for fluctuations in the of market value of Counterparty Credit Risks. These have to be computed from market implied estimates of the individual counterparty's default risks. Also, effective in 2013, the IFRS 13 Fair Value Measurement issued by the International Accounting Standard Board (IASB, 2011)   
requires financial entities to report the market values for their derivative positions, including the market-implied assessment of Counterparty Credit Risks. 

The calculation of both the CVA and of the associated regulatory capital requires the calibration of a counterparty's term structure of risk-neutral default probabilities to liquid Credit Default Swap (or CDS) quotes associated with the counterparty. However, as highlighted in a European Banking Authority's survey (EBA, 2015), %\cite{EBAreport})   
among the Internal Model Method (or IMM) compliant banks, over 75\% of the counterparties do not have liquid CDS quotes. We refer to the problem of assessing the default-risk of counterparties without liquid CDS quotes as the \textit{Shortage of Liquidity problem}, and to any method to construct the missing quotes on the basis of available market data as a \textit{CDS Rate Construction Method}.   
   
One recommended solution-method for this Shortage of Liquidity problem is to map a counterparty lacking liquidly quoted CDS spreads to one for which a liquid market for such quotes exists, and which, according to criteria based on financial market data, best resembles the non-quoted party. The CDS data of the liquid counterparty, which is called a {\it proxy} of the non-liquid counterparty, can subsequently be used to obtain estimates for the risk-neutral default probabilities of the latter. Following practitioners' usage, we will refer to this as a \textit{CDS Proxy Method}. We will refer to the counterparties having liquid CDS quotes as \textit{Observables} and those without liquid quotes as \textit{Nonobservables}. We note in passing that often, in practice, if, for a given day, the number of a counterparty's CDS quotes from available data vendors such as MarkIT\texttrademark (or others) falls below a certain threshold, the counterparty is deemed to be illiquid.  

As we will see below, current methods for constructing CDS proxy rates follow a different route, in that they seek to directly construct the missing rates, based on Region, Sector and Ratings data. They treat all counterparties in a given Region, Sector and Ratings bucket homogeneously and therefore typically fail to pick up counterpart-specific default risk, thereby failing some of the European Banking Authority's criteria for a sound CDS proxy-rate construction method. In this paper we will, amongst other things, enlarge the set of financial market data used for CDS Proxy construction to include both stock and options market data. We also replace ratings by estimated default probabilities. It might have been considered natural to also include corporate bond spreads in this list, but we decided not to do so because of the relative lack of liquidity in the corporate bond market: Longstaff {\it et al.} (2005) %\cite{Longstaffliquidity}   
concluded from a comprehensive empirical study of CDS spreads and the corresponding bond spreads that there exists a significant non-default component in corporate bond spreads due to illiquidity, as  measured by the bid-ask spreads of corporate bonds. As highlighted by Gregory (2015), %\cite{Gregory},   
the vast majority of the counterparties of banks do not have liquid bond issues and the liquidity of bonds is generally considered to be poorer than that of the CDS contracts associated with the same counterparties. Therefore, a bond-spread based CDS Proxy Method will not be particularly helpful for solving the Shortage of Liquidity problem. That being said, it would be easy to include such bond spread variables into any of the CDS Proxy methods we introduce and examine in this paper.

Another contribution of this paper is that we go beyond the traditional regression approach commonly used in Finance, and construct our CDS Proxies using classification algorithms which were developed by the Machine Learning community, algorithms whose efficiency we tested and whose relative performances we rated. As discussed below, there is little published research on the Shortage of Liquidity problem and its solution. One of the motivations of this paper is to address what seems to be an important gap in the literature, by providing and examining alternatives to the few currently existing CDS Rate Construction Methods.        
   
\subsection{Regulators' Criteria for Sound CDS Proxy Methods}\label{Bucketing}      
According to publications by the Basel Committee for Banking Supervision, (BCBS, 2010) and (BCBS, 2015), and the criteria specified by the European Banking Authority (EBA, 2013), any CDS Proxy Method should, at a minimum, satisfy the following criteria:    
\begin{enumerate}
\item The CDS Proxy Method has to be based on an algorithm that discriminates at least three types of variables: Credit Quality (e.g., rating), Industry Sector and Region (BCBS, 2015). 
\item Both the Observable Counterparties used to construct a CDS proxy spread for a given Nonobservable and the Nonobservable itself have to come from the same Region, Sector and Credit Quality group (BCBS, 2015). 
\item According to (EBA, 2013), 
the proxy spread must reflect available credit default swap spreads. In addition, \textbf{\textit{the appropriateness of a proxy spread should be determined by its volatility and not by its level}}. This criterion highlights the regulators' requirement that CDS proxy rates should include counterparty-specific components of counterparty default risk, which are not adequately measured by purely level-based measures such as an average or median of a group of CDS spreads or region- or industry-based CDS indices.   
\end{enumerate}
We next take a look at existing CDS Proxy Methods. A survey of the publicly available literature showed that the following two types of CDS Proxy Methods are currently used by the finance industry:
 
\begin{itemize}
\item  The \textit{Credit Curve Mapping Approach} described in Gregory (2015), which simply creates Region/Sector/Rating buckets consisting of single-name CDS contracts with underlying reference entities from the same regions and sectors and with the same ratings, and then proxies the CDS rates of a Nonobservable from a given Region/Sector/Rating bucket by the mean or median of the spreads of the single-name CDS rates within that  bucket. This approach clearly satisfies criteria \#1 and \#2 but assumes the counterparty default risks for all counterparties coming from the same Region, Sector and Rating bucket to be homogeneous and ignores the idiosyncratic default risk specific to individual counterparties. In particular, using bucket average or median as Proxy CDS rate for a Nonobservable ignores the CDS-spread volatility across counterparties within the bucket, thereby failing to meet criteria \#3.   
   
\item The \textit{Cross-sectional Regression Approach} of Chourdakis {\it et al.} (2013): %\cite{Nomura}:   
this is another popular CDS Proxy Method which assumes that an observable counterparty $i $'s CDS rate $S_i $ (for a CDS contract with given maturity and payment dates) can be explained by the following log-linear regression equation: 
\begin{align}\label{crosssection}
%S^{proxy}_i= \beta_{global} * \beta_{region} * \beta_{sector} * \beta_{rating}*\beta_{seniority} \nonumber \\ \text{  alternatively,   }
\log (S_i)=\beta_0+\sum_{m = 1 }^{\# {\rm Regions } } \beta ^R_m I^R _{i,m } +\sum_{m = 1 }^{\# \rm{ Sectors } } \beta ^S _m I^S _{i , m } +\sum_{m = 1 }^{\# \rm{ ratings } } \beta^r _m I^r _{i,m } + \sum_{m = 1 }^{\# {\rm seniorities} } \beta ^s _{m } I^s _{i, m } +\epsilon_i ,   
\end{align}  
where the $I $'s s are dummy or indicator variables for, respectively, sector, region, rating class and seniority,   
as specified in the CDS contract. The regression-coefficients can be estimated by Ordinary Least Squares, with $\epsilon _i $  representing the estimation error. Once estimated, the regression can be used to predict the rate of a CDS-contract of a Nonobservable in a given region and sector with a given rating and seniority.    The Cross-sectional Regression Approach goes beyond the Curve Mapping Approach in that it provides a linkage between CDS Rates and the above five explanatory variables, which goes beyond simply taking bucket means. However, like the Curve Mapping Approach, its predictions still treat the market-implied default risks for counterparties within the same Region, Sector, Rating and Seniority bucket homogeneously and as such ignores counterparty-specific risks. The Cross-sectional Regression Approach proxies the counterparty's CDS spread by an expected value coming from a regression; it provides the level of a CDS spread but ignores the CDS-spread volatility among individual counterparties and therefore does not satisfy criteria \#3 either.      
\end{itemize}   
   
As we have seen, neither the Curve Mapping Approach nor the Cross-sectional Regression Approach satisfies EBA's criterion \#3. Furthermore, we have not encountered any out-of-sample performance tests for either model to assess the reliability of their predictions. Such tests are a critical ingredient of any statistical modelling approach. The different Machine Learning-based CDS Proxy Methods we introduce and investigate in this paper do take idiosyncratic counterparty default risk into account. Moreover, we performed extensive out-of-sample tests for each of the methods we investigated using Stratified Cross Validation. We briefly describe what we believe are the main contributions of this paper to the existing literature.   
   
\subsection{Contributions of the Present Paper}
To tackle the Shortage of Liquidity problem, we apply Machine Learning Techniques, and more specifically, Classification Techniques, using the best known Classifier families in the Machine Learning area: Discriminant Analysis (Linear and Quadratic), Na\"ive Bayesian, $k $-Nearest Neighbourhood, Logistic Regression, Support Vector Machines, Neural Networks, Decision Trees and Bagged Trees. We call this approach the Machine Learning based CDS Proxy Method. 
To the best of our knowledge, this paper represents the first research in the public domain on applications of Machine Learning techniques to the Shortage of Liquidity Problem for CDS rates.      
   
Furthermore, given the wide range of available Classifiers and their parametrisations choices, the question of comparing the performances of different classifiers naturally arises. We have carried out a detailed classifier performance comparison study across the different Classifier families (referred to as Cross-classifier Comparison below) and also within each Classifier family (referred to as Intra-classifier Comparison). As far as we are aware, this is the first empirical comparison of Machine Learning Classifiers based entirely on financial market data. To clarify this point, we briefly review some of the existing Classifier Comparison literature. Readers who are unfamiliar with the Machine Learning-terminology used in this section may refer to Section \ref{sectionclassmodel} for clarification.  

STATLOG by King {\it at al.} (1995) is probably the best known empirical classifier comparison study for a comprehensive list of classifiers. One conclusion drawn from that study is that classifier performance can vary greatly depending on the type of dataset used, such as sector-type: Medical Sciences, Pharmaceutical Industries and others, and that researchers should rank classifiers' performance for the specific type of data set which is the subject of their study. As mentioned, we believe ours to be the first such study for financial market data.   

Delgado and Amorim (2014) is a more recent contribution to the classifier comparison literature. It compared 179 classifiers from 17 families based on 121 datasets, using \textit{Maximum Accuracy} as the criterion to identify the top performing classifier families.  As emerged from their study, these were, in descending order of performance, the Random Forest (an example of a so-called Ensemble Classifier), the Support Vector Machine, more specifically with Gaussian or Polynomial Kernels, and the Neural Network. All of these will be presented in some detail in Section \ref{sectionclassmodel} below, except for the Random Forest algorithm, which in our study we replaced with another Ensemble Classifier, the Bagged Tree.   

For our \textit{cross-classifier} performance comparison 
we have rank-ordered classifiers according to Expected Accuracy as estimated by $K$-fold Stratified Cross Validation with $K=10 $ (a choice which we show  to be empirically justified). Existing literature on classifier performance typically compare 
different families of classifier algorithms without examining in much detail performance variations within each of the families or the effects of feature-variable selection. As discussed in Section \ref{empsection} below, our empirical results show that the latter can contribute significantly to variations in classifier performance, both within and across classifier families. The inclusion of feature-variable selection in cross-classifier comparison is an original contribution of our study. Furthermore, our study has examined the impact on classifier performance of correlations amongst feature-variables. Here it is to be noted that, in contrast to many of the data sets used in previous comparison studies, financial market data can, and typically will, be strongly correlated, especially in periods of financial stress such as the one on which we based our study (the 100 days leading up to Lehman's 2008 bankruptcy). We believe ours to be one of the first research efforts to understand the impact of multicollinearity on classification performance in the specific context of financial markets. Depending on the classifier, we found that this impact can either be negligible (for most Classifier families) or negative, for Na\" ive Bayes and for some of the LDA and QDA classifiers.      
   
We have compared a total of 156 classifiers from the eight main families of classifier algorithms mentioned above, with different choices of parameters (which can be functional as well as numerical) and different choices of feature variables, for the construction of CDS Proxies on the basis of financial data from the 100 days up to Lehman' bankruptcy. Our Cross-classifier comparison shows that for the construction of these Proxies, the three top-performing classifiers are the Neural Network, the Support Vector Machine and the Bagged Tree. A graphical summary of our results is given in Figure \ref{FigureAllClassifierAll}. This performance result is broadly consistent with that of both King {\it at al.} (1995) and Delgado and Amorim (2014). As part of our \textit{Intra-classifier} performance comparison, we have compared classifiers within a single family with different parameterisations and different feature variables and found that this performance typically varies greatly, though for certain Classifier families, such as the Neural Networks, it can also be relatively stable. The empirical results are further discussed in Section \ref{empsection}, on the basis of graphs and tables which are collected in Appendix \ref{sectionindividaulclass}.   
   
Although the different Classifier algorithms produced by the Machine Learning community can be used as so many `black boxes' (which is arguably one of their advantages, from a user's perspective), we believe that is also important to have at least a basic understanding of these algorithms, for their correct use and interpretation, and also to understand their limitations. For this reason we have included a section where we present each of the eight Classifier families we have used, in the specific context of the problem we are addressing here.

\subsection{Structure of the Paper}
The rest of the paper is organized into three Sections plus two Appendices:   
\begin{itemize}
\item In Section \ref{sectionclassmodel}, we give a brief introduction to Machine Learning classification with a description of each of the eight families of classifier algorithms we have used, alongside an illustrative example in the framework of our CDS Proxy Construction problem.   
   
\item In Section \ref{empsection}, we present the results of our cross-classifier and intra-classifier performance comparison study for the construction of CDS Proxies.   
   
\item Section \ref{SectionConclusion} presents our conclusions and provides some directions for future research.   
   
\item Appendix \ref{sixmodels}, finally, gives a detailed description of the six feature selections which are at the basis of our Proxy constructions, and describes the data which we have used, while Appendix \ref{sectionindividaulclass} contains the different graphs and tables related to individual classifier performances needed for the discussion of section 3.      

\end{itemize}
\section{Classification Models}\label{sectionclassmodel}   
    
Compared with traditional statistical Regression methods, \textit{Machine Learning} and \textit{Classification Techniques} are not as well-known in the Finance industry and, for the moment at least, less used. This section introduces the basic concepts of Machine Learning and presents the eight classification algorithms which we used for our CDS Proxy construction. As a general reference, Hastie {\it et al.} (2009) provides an excellent introduction to Classification and to Machine Learning in general. Although our paper focuses on %presenting the classification concepts and models in the context of   
the CDS Proxy Problem, its general approach can easily be adapted for the construction of proxies for other financial variables for which no or not enough market data are available.    
   
The general aim of Machine Learning is to devise computerised algorithms which predict the value of one or more \textit{response variables} on a population each of whose members is characterised by a vector-valued \textit{feature variable} with values in some finite dimensional vector space $\mathbb{R}^d . $ For  our CDS Proxy Problem, the population in question would consist of a collection of counterparties, each of which, for the purpose of constructing the proxy, would be characterised by a number of discrete variables such as region, sector or rating class, and continuous variables such as historical and implied equity volatilities and (objective) Probabilities of Default or PDs, all over different time-horizons. We will sometimes add the 5-year CDS rate to this list, which is the most quoted rates, and therefore may be available for counterparties for which liquid quotes for the other maturities are missing.   
  
The construction of the feature space should be based on its statistical relevance and on the economic meaning of the feature variables. For the former, formal statistical analysis can be conducted to rank-order the explanatory power of each of the feature variables regarding the response variable. As regards their economic relevance, one can apply business judgement, and also use available research on explanatory variables for CDS rates. A further important issue for us was that the features used should be liquidly quoted for the nonobservable counter parties as well as the observable ones. In our study, we have selected six sets of feature variables on the basis of which we constructed our CDS proxies, and we refer to Appendix \ref{sixmodels} for their description and for some further comments on how and why we selected them as we did. Our selection is not meant to be prescriptive, and in practice a user might prefer other features.   
      
The relevant response variable could be a CDS rate of a given maturity, as in the two existing CDS Proxy Methods mentioned in section 1.2, or it could be a class label, such as the name of an observable counterparty.  This already shows that the response variable can either be \textit{continuous}, in which case we speak of a \textit{Regression Problem}, or \textit{discrete}, in which case we are dealing with a \textit{Classification Problem}. In this paper we will be uniquely concerned with the latter. 
     
Unlike Regression-based approaches, the Classification-based CDS Proxy Methods we investigate in this paper do not attempt to predict the CDS rates of the different maturities. Instead, they associate an observable counterparty to a non-observable one based on a set of (financial) features which can be observed for both. The, liquidly quoted, CDS rates of the former can then be used for managing the default risk of the latter, such as the computation of CVAs and of CVA reserve capital. As compared to regression, a Classification-based CDS Proxy method has the advantage of ultimately only using market-quoted CDS rates, which therefore, in principle at least,  are free of arbitrage. By contrast, using a straightforward regression for each of the different maturities for which quotes are needed risks introducing spurious arbitrage opportunities across the different maturities: see Brummelhuis and Luo (2017). %\cite{Brum_Luo}   
It in fact turns out to be remarkably difficult to precisely characterise arbitrage-free CDS term structures even for a basic reduced-form credit risk model (though simple criteria for circumventing "evident" arbitrages can be given). On the other hand, such a characterisation would seem to be a necessary pre-requisite for an guaranteed arbitrage-free regression approach. We make a number of further observations.   

\begin{enumerate}
\item Depending on the type of classifier, the strength of the association between an observable and a nonobservable counterparty can be defined and analysed statistically. For example, Linear Discriminant Analysis which, going back to R. A. Fisher in 1936, is one of the oldest classifier algorithms known, estimates the posterior probability for a nonobservable to be classified into any of the observables under consideration, given the observed values of the feature variables. It then chooses as CDS Proxy that observable for which this posterior probability is maximal: see subsection\ref{ldasection} below for details. 
   
\item The different classifiers will be calibrated on Training Sets, and their performances evaluated by a statistical procedure know as {\it $K $-fold Stratified Cross Validation}: see section \ref{SectionKFold} below for a description. Depending on the strength of the statistical association and the results of the cross-validation, one can then argue that the nonobservable will, in its default risk profile, resemble the observable to which has been classified.   
   
\item When applying our Classification-based CDS Proxy methods, we will take the observables and nonobservables from a same region/sector bucket, but we won't bucket by rating. Instead, we will use the probabilities of defaults (PD) over different time-horizons as feature variables. This means that in practice our classification will be based on the continuous feature variables only. The different PDs might be provided by a Credit Rating Agency or be a bank's own internal ratings.   
   
\item We ultimately take as missing CDS rates for a Nonobservable counterparty the, market-quoted, CDS rates of the Observable to which it is most strongly associated by the chosen Machine Learning algorithm, on the basis of the values of its feature variables (where the precise meaning of "association" will depend on the algorithm which is used). In view of the previous point, these proxy rates will automatically reflect Region and Sector risk, and thereby satisfy the regulators' criteria \#1  and  \#2 mentioned in subsection \ref{Bucketing} above. Our approach also addresses the regulators' criterion \#3: the proxy rates will be naturally volatile, as market-quoted, rates of the selected Observable. Furthermore, they will also reflect the Nonobservable's own counterparty-specific default risk since, as market conditions evolve, the non-observable may be classified to different observables depending on the evolution of its feature variables.   
\end{enumerate}   
   
To give a more formal description of a Classification algorithm, let $\{ 1, \ldots , N \} $ be the set of Observables, which we therefore label by natural numbers, and let $\mathbb{R }^d $ be the chosen feature space. A typical feature vector will be denoted by a boldface $\mathbf{x } = (x_1 , \ldots , x_d ) . $ As mentioned, for us the components of the feature vector will be financial variables such as as historical or implied volatilities and/or estimated PDs, over different time horizons. Our Classification problem is to construct a map $\widehat{y } : \mathbb{R }^d \to \{ 1 , \ldots , N \} $ based on a certain {\it training set} of data,   
\begin{equation}  \label{EqTrainingSet} 
D^T = \{ (\mathbf{x }_i , y_i ) : \mathbf{x }_i \in \mathbb{R }^d , y_i \in \{ 1, \ldots , N \} , \,  \ i = 1 , \ldots ,n \, \} ,   
\end{equation}   
corresponding to data on Observable counterparties: $\mathbf{x }_i $ is an observed feature vector of counterparty $y_i . $ Each of the Machine Learning algorithms we will use is, mathematically speaking, a recipe for the construction of a map $F_{\theta } : \mathbb{R }^d \to \{1 , \dots , N \} $, with $\theta $ a vector of parameters, and our maps $\widehat{y } $ will be of the form:   
\begin{equation}  
\widehat{y } (\mathbf{x } ) = F_{\widehat{\theta } }(\mathbf{x})
\end{equation}   
where the parameters $\widehat{\theta } $ will be ''learned'' from the training set $D^T $, usually by maximising some performance criterion or minimising some error, depending on the type of Learning algorithm used. The parameters $\theta $ can be numerical, such as the $k$ in the $k$-Nearest Neighbour method, the tree size constraint in the Decision Tree algorithm or the number of learning cycles for the bagged Tree, but also functional, such as the kernel function in the Na\"ive Bayesian method, the kernel function of a Support Vector Machine, or the activation function of a Neural Network. Where possible, we have optimised numerical parameters by cross-validation.   
   
We note in passing that the constructed classifier map $\widehat{y } $ will in general of course strongly depend on the training set and a more complete , though also more cumbersome, notation such as $\widehat{y }_{D^T } $ (or $\widehat{\theta } _D $ for the optimal parameters) would have indicated this dependence; we will however leave it as implicitly understood. In practice there would also be the question of how often we would need to update these training sets: this would amongst other things depend on the speed with which the classification algorithm "learns" - determines the 

Following Machine Learning literature such as Delgado and Amorim (2014), we refer to the classification algorithms that are based on the same methodology as a \textit{Classifier Family}. Within each Classifier Family, we refer to classification algorithms that differ from each other by their parameterisations, including the dimensionality of the feature vectors, as the individual  \textit{Classifiers}. As shown in later sections, exploring parameterisation choices not only serves to choose the best classifier based on classification performance, it is also helpful to explain performance variations across classifiers.    
 
Table \ref{classifiersfig} lists the 156 Classifiers from the 8 Classifier Families we investigated in this paper. The first column contains the labels for the Classifiers which will be used in the rest of paper, including the Appendices. The second column contains a brief description of each of the Classifiers and the headings ''FS1-FS6'' in the third column refers to the 6 different feature variable selections we have used and which are detailed in Appendix \ref{sixmodels}.   
   
\begin{table}
\caption{156 Classifiers under eight most popular Classifier Families;''FS'' stands for ''Feature Selection''}\label{classifiersfig} 
\centering   
%\hspace*{-1.5cm} 
\includegraphics[scale=1.0]{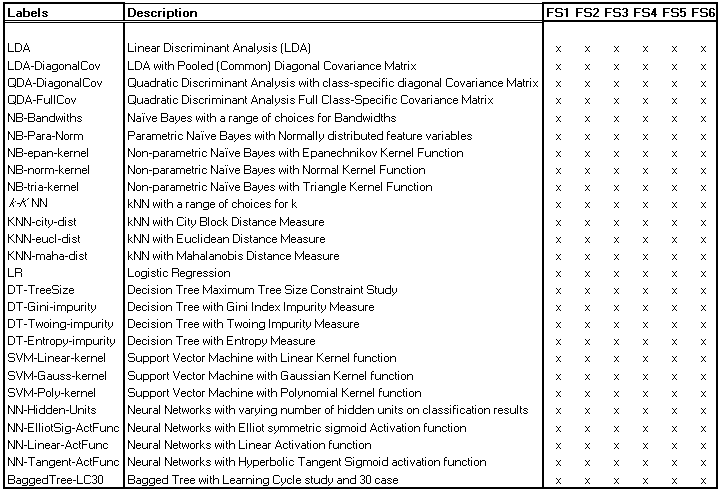} 
\end{table}

In the remainder of this section, we introduce the eight most popular Classifier Families of Machine Learning (Wu {\it et al.}, 2008) , in the specific context of the CDS Proxy problem, together with some illustrative examples. At the end of the section we present the statistical procedure we used for cross validation and Classifier selection. We next turn to a description of each, with illustrative examples.   
\medskip

\subsection{Linear Gaussian Discriminant Analysis}\label{ldasection}   
Linear Discriminant Analysis or LDA, as well as the closely related Quadratic Discriminant Analysis (QDA) and Na\"ive Bayesian Method (NB) which we will discuss below, are based on Bayes' rule.  
These methods interpret the training set $D^T = \{ (\mathbf{x }_i , y_i ) : i = 1, \ldots , n \} $ %with $\mathbf{x }_i \in \mathbb{R }^p $ is a feature vector of the observable counter-party $y_i \in \{ 1, \ldots , N \} $   
as a sample of a pair of random variables $(\mathbf{X } , Y ) $, and estimate the posterior probability density that a feature vector $\mathbf{x } $ is classified into the class (for us, counterparty) $j $ by Bayes' formula,   
\begin{align} \label{eqgaussian}
\mathbb{P } ( Y = j \mid \textbf{X } = \mathbf{x })=\frac{\mathbb{P }(\textbf{X } = \mathbf{x } \mid Y = j ) \, \mathbb{P }(\textbf{Y}=j)}{\mathbb{P } (\textbf{X } ) } ,   \nonumber %\\
%where\ P(\textbf{X})=\sum_{k=1}^M P(\textbf{X}\mid \textbf{Y}=k)P(\textbf{Y}=k)
\end{align}   
where $\mathbb{P } (\mathbf{X } = \mathbf{x } | Y = j ) $ is a prior probability density on the feature vectors belonging to the class $j $, which we will also call the {\it class-conditional density}, and where $\mathbb{P }(\textbf{X})=\sum_{k=1}^N \mathbb{P }(\textbf{X } \mid Y=k ) \mathbb{P }(Y = k) . $ To simplify notations we will often write $\pi _j := \mathbb{P } (Y = j ) $ for the, unconditional, probability of membership of class $j $, $\mathbf{P } (\mathbf{x } | j ) $ for $\mathbb{P } ( \textbf{X } = \mathbf{x } \mid Y = j ) $ and $\mathbf{P } (j \mid \mathbf{x } ) $ for $\mathbb{P } ( Y = j | \mathbf{X } = \mathbf{x } ) . $

The LDA, QDA and NB methods differ in their assumptions on the class-conditional densities. For the first two these densities are assumed to be Gaussian. One can generalize the LDA and QDA methodologies to include non-Gaussian densities, e.g. by using elliptical distributions, but the resulting decision boundaries (defined below) would no longer have the characteristic of being the linear or quadratic hyper surfaces which give their name to LDA and QDA.   We refer to Hastie {\it et al.} (2009) for a general introduction to Bayesian Classifications based on Bayes' formula.      
   
\subsubsection{The LDA algorithm}      
   
Linear Discriminant Analysis defines the class $j $ conditional density function to be      
\begin{equation}
f_j (\mathbf{x } ) := \mathbb{P } (\mathbf{X } = \mathbf{x } | Y = j ) :=   
(2\pi)^{-\frac{d}{2}} |V|^{-\frac{1}{2}} 
%\exp \{ 
e^{-\frac{1}{2} ( \mathbf{x } - \boldsymbol{\mu }_j )^T V^{-1} ( \mathbf{x } - \boldsymbol{\mu }_j )} ,   
\end{equation}
where $\boldsymbol{\mu }_j $ represents the mean of the feature vectors associated to the $j $-th class and where $V $ is a variance-covariance matrix with non-zero determinant $|V | $ and inverse $V^{-1 } . $ Observe that we are taking {\it class-specific} means but a {\it non-class-specific} variance-covariance matrix. Taking the latter class-specific also leads to QDA, which will be discussed in subsection 2.2 below.   
   
As a natural choice for $\boldsymbol{\mu }_j $ and  $V $ we can, for a given training data set $D^T = \{ \mathbf{x }_i , y_i ) : i = 1 , \ldots , n \} $, take for $\boldsymbol{\mu }_j $ the sample mean of the feature vectors associated to class $j $,   
$$   
\boldsymbol{\mu }_j = \frac{1 }{n_j } \sum _{i : y_i = j } \mathbf{x }_i ,   
$$   
where $n_j := \# \{ (\mathbf{x }_i , y_i ) : y_i = j \} $ is the number of data points in class $j $, and for $V $ be the sample variance-covariance matrix of $\{ \mathbf{x }_i : i = 1, \ldots , n \} $, the set of all features vector of the training sample. Alternatively, we can estimate these parameters by Maximum Likelihood; we note that for normal distributions the maximum likelihood estimators for mean and variance-covariance are of course asymptotically equal to their sample values, but that here the unconditional distribution of $\mathbf{x } $ will not be normal, but a normal mixture.

We finally need the prior probabilities $\pi _j $ for membership of class $j  $; again, several choices are possible. We have simply taken the empirical probability implied by the training set by putting $\pi _j = n_j / n $, where we recall that $n = \# D^T $ is the number of data points, but, alternatively, one could impose a Bayesian-style uniform probability $\pi _j = 1 / N $ (which would be natural as an initial a-priori probability if one would use Bayesian estimation to train the algorithm).   
Finally, one can estimate the $\pi _j $'s by Maximum Likelihood, simultaneously with the other parameters above.      
\medskip   
   
Once calibrated or "learned", a new feature vector $\mathbf{x } $ is classified as follows: the log-likelihood ratio that $\mathbf{x } $ belongs to class $j $ as opposed to class $l $ can be expressed as:
\begin{align}\label{LDAloglh}
\mathcal{L}_{j, l } (\mathbf{x } ) := \log \left( \frac{ f_j (\mathbf{x } ) \pi_j / \mathbb{P }(\mathbf{x } ) }{ f_l(\mathbf{x }) \pi_l / \mathbb{P }(\mathbf{x } ) } \right) = \log\ (\pi_j / \pi_l ) + \log f_j (\mathbf{x } )  - \log f_l (\mathbf{x } ) \nonumber \\
= \log (\pi_j / \pi_l ) - \frac{1}{2} \boldsymbol{\mu }_j^T V^{-1} \boldsymbol{\mu }_j + \frac{1}{2} \boldsymbol{\mu }_l^TV^{-1} \boldsymbol{\mu }_l + \mathbf{x }^T V^{-1} (\boldsymbol{\mu }_j - \boldsymbol{\mu }_l ) .   
\end{align}   
This is an affine function %$\beta ^T \mathbf{x } + \beta _0 $ of $\mathbf{x } $ with coefficients which can be read off from the above expression,   
of $\mathbf{x } $ and the decision boundary, which is obtained by setting $\mathcal{L}_{j, l } (\mathbf{x } ) $ equal to 0, is therefore a hyperplane in $\mathbb{R }^d . $ If we define the discriminant function for class $j $ by   
\begin{equation} \label{decisionlda}
d_j(\mathbf{x } )=\mathbf{x }^T V^{-1} \boldsymbol{\mu_j } -\frac{1}{2} \boldsymbol{\mu_j }^T V^{-1} \boldsymbol{\mu_j } + \log \pi _j ,    
\end{equation}   
then $\mathcal{L}_{j, l } (\mathbf{x } ) > 0 $ iff $d_j (\mathbf{x } )>d_l (\mathbf{x } ) $, in which case the feature vector $\mathbf{x } $ is classified to observable $j $ as opposed to $l . $ This is the LDA criticism 
for classification into two classes. To generalise this to multi-class classification, we classify a feature vector $\mathbf{x } $ into the class $j $ to which it has the strongest association as measured by the discriminant functions (\ref{decisionlda}): our optimal decision rule therefore becomes   
\begin{equation}\label{eqlda1}
\widehat{y }(\mathbf{x } ) =\arg\max_j d_j (\mathbf{x } ) .   
\end{equation}
This is also known as the {\it Maximum A Posteriori} or MAP decision rule; the same, or a similar, type of decision rule will be used by other classifier families below.   
   
\subsubsection{An Illustrative Example}\label{running_DA_example}   
   
\begin{figure}[h]
\caption{Linear Discriminant Analysis}\label{LDAfigure}
\centering
\includegraphics[scale=0.6]{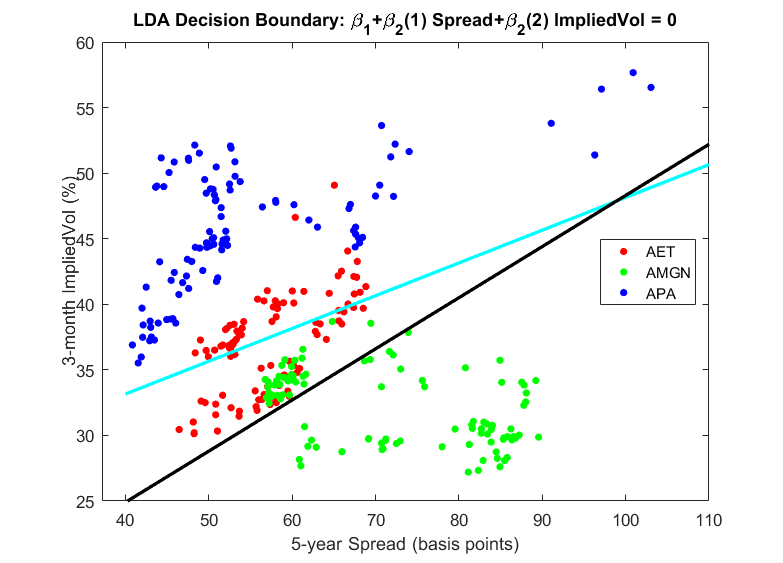} 
\end{figure}

To illustrate the two DA algorithms and NB we use the example of three counterparties called AET, APA and AMGN, for which we have observed two-dimensional feature vectors $\mathbf{x } = (s,\sigma _{3m } ^{imp } ) $ representing the 5-year CDS rate and the 3-month implied volatility. In Figure \ref{LDAfigure} the observed feature vectors of APA are plotted in blue, and those of AET and AMGN in red and green, respectively. The cyan line is the resulting LDA decision boundary  discriminating between APA and AMGN while the dark line discriminates between AET and AMGN. We have omitted the third decision boundary for better lisibility of the graph. Clearly, the LDA does a better job discriminating between APA and AMGN than between AET and AMGN. To classify a nonobservable counterparty, one first computes the "scoring functions" (\ref{decisionlda}), with  $j = 1, 2, 3 $ corresponding to the three observables, APA,  AMGN and AET, respectively, and associates the counterparty to that $j $ for which the score is maximal (and for example to the smallest $j $ for which the score is maximal in the exceptional cases there are several such $j $).   

As indicated in Table \ref{classifiersfig}, we have investigated two types of LDA classifiers with two different parameterisation choices for the pooled variance-covariance matrix $V$, one imposing a diagonal $V $ with the sample variances on the diagonal, and the other with the full (empirical) variance-covariance matrix $V $, and tested these with the six feature-variable selections of Appendix  \ref{sixmodels}. We refer to section \ref{empsection} for the comparison of LDA's performance with that of the other Classifier families, and to Appendix \ref{sectionindividaulclass} for the intra-class performance comparison.        
   
\subsection{Quadratic Discriminant Analysis}\label{qdasection}
LDA assumes that the variance-covariance matrix of the class-conditional probability distributions is independent of the class.   
By contrast, in Quadratic Discriminant Analysis (QDA) we allow a class-specific covariance matrix $V_j $ for each of the classes $j . $ As a result, we will get quadratic decision boundaries instead of linear ones.   
\subsubsection{The QDA algorithm}   
   
Under the Gaussian assumption, the class conditional density functions for QDA are take to be:
\begin{equation}\label{qdarationale}
f_j (\mathbf{x } )= \mathbb{P }(\mathbf{X } = \mathbf{x } \mid Y = j ) = (2\pi)
^{-\frac{d }{2 } } |V_j|^{-\frac{1}{2} } e^{-\frac{1}{2} (\mathbf{x } - \boldsymbol{\mu }_j )^T V_j^{-1} (\mathbf{x } - \boldsymbol{\mu } _j ) } ,   
\end{equation}
where now $V_j$ is a class-specific variance-covariance matrix and $\boldsymbol{\mu }_j $, as before, a class-specific mean. As for LDA, there are several options for calibrating the model; we simply took the sample mean and variance-covariance matrix of the set $\{ \mathbf{x }_i : y_i = j \} . $ %or we can learn these matrices from the training set through some optimization procedure, together or not with the $\boldsymbol{\mu }_j $ and $\pi _j $'s.         

Comparing log-likelihoods of class memberships as we did for LDA now leads to quadratic discriminant functions $d_j $ given by   
\begin{equation}\label{qdadecision}
d_j (x )=-\frac{1}{2} \log |V_j |-\frac{1}{2} (x-\mu_j )^T V_j^{-1} (x-\mu_j ) + \log\pi_j .   
\end{equation}
A feature vector $\mathbf{x } $ will be classified to class $j $ rather than $l $ if $d_j (\mathbf{x } ) > d_j (\mathbf{x }_l ) $, and the decision boundaries $\{ \mathbf{x } : d_j (\mathbf{x } ) = d_l (\mathbf{x }_l ) \} $ will now be quadratic. The Decision Rule for multi-class classification under QDA is again the MAP rule (\ref{eqlda1}), but with the new scoring function (\ref{qdadecision}). If $V_j = V_l $ for all $l $ and $j $, then QDA reduces to LDA.   
\medskip

\subsubsection{An Illustrative Example for Quadratic Discriminant Analysis}
\begin{figure}[h]
\caption{An Example for Quadratic Discriminant Analysis}\label{QDAgraph}
%\hspace*{-1.5cm}    
\centering
\includegraphics[scale=0.65]{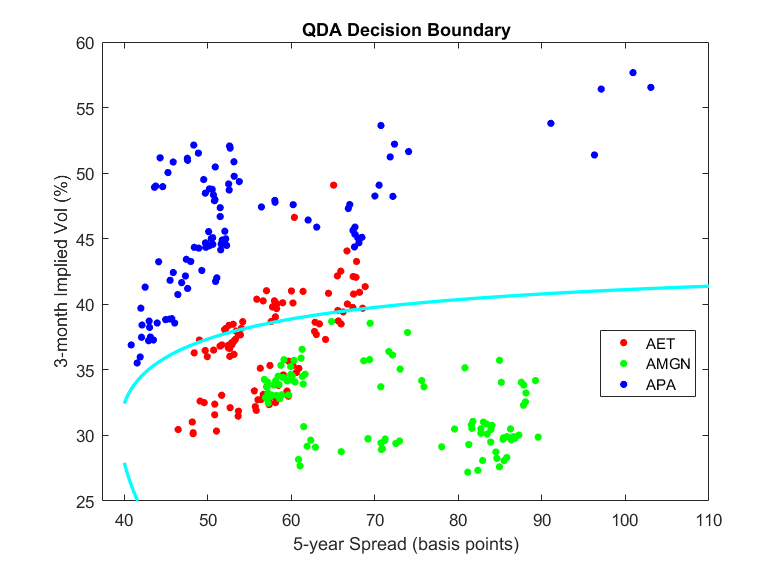} 
\end{figure}

The cyan curve in Figure \ref{QDAgraph} depicts the quadratic decision boundary between counterparties APA and AMGN of example \ref{running_DA_example} as found by the QDA algorithm, where for the purpose of better presentation we only show one of the three decision boundaries.   
   
As indicated in Table \ref{classifiersfig}, we have investigated two QDA classifiers with, as for LDA, two parametrisation choices for the covariance matrices $V_j $, diagonal and full, for the six different Feature Variable Selections. Cross-classifier and intra-classifier comparison results can be found in Section \ref{empsection} and Appendix \ref{sectionindividaulclass}. In particular, Figure \ref{DAAccuracy} shows that for our CDS Proxy problem, QDA with full variance-covariance matrices outperforms the other DA algorithms across all six Feature Selections, where the difference in performance can be up to around 20\% in terms of accuracy rates.   
   
\subsection{Na\"ive Bayes Classifiers}\label{sectionNB}      
As for LDA and QDA, Na\"ive Bayes Classifiers calculate the posterior probabilities $\mathbb{P } ( j \mid \mathbf{x } ) $ using Bayes' formula, but make the strong additional assumption that, within each class, the components of the feature variables act as independent random variables: given that $Y = j $, the components $X_ {\nu } $ of $\mathbf{X } $ are independent, $\nu = 1 , \ldots , d . $ In other words, the individual features are assumed to be conditionally independent, given the class to which they belong. As a consequence of this  \textit{Class Conditional Independence assumption}, Na\"ive Bayes reduces the estimation of the multivariate probability density $\mathbb{P }(\textbf{x} \mid j ) $ to that of the $d $ univariate probability densities,   
$$   
f_{j , \nu } (x ) := \mathbb{P } (X_{\nu } = x | Y = j ) , \ \ \nu = 1, \ldots , d ,      
$$   
with the class-conditional densities being simply given by their product.   

\subsubsection{The NB decision algorithm}   
   
Under the Class Conditional Independence assumption, the class $j$ conditional density is     
\begin{equation} \label{naiveprod}
f_j (\boldsymbol{x})=\prod_{\nu = 1}^d f_{j, \nu } (x_{\nu } ) .   
\end{equation}
The log-likelihood ratio (\ref{LDAloglh}) can be evaluated as   
\begin{align}\label{eqNaiveBayes}
\mathcal{L}_{j, l } (\mathbf{x } ) = \log \left(\frac{\pi_j f_j (\boldsymbol{x } ) }{\pi_l f_l (\boldsymbol{x})} \right) %=log \frac{\pi_j \prod_{d=1}^p f_{j,d} (x_d ) }{\pi_l \prod_{d=1}^p f_{l,d} (x_l ) }\nonumber \\
= \log \frac{\pi_j}{\pi_l} +\sum_{\nu =1 }^d \left( \log f_{j, \nu }  (x_{\nu } ) -  \log f_{l,\nu } (x_{\nu } \right) ,   
\end{align}   
with the decision boundaries again being obtained by setting $\mathcal{L }_{ij } (\boldsymbol{x } ) $ equal to 0. The discriminant functions of Na\"ive Bayes are therefore   
\begin{equation}\label{naivebayesdiscriminant}
d_j(\boldsymbol{x } ) = \log \pi_j +\sum_{\nu = 1}^d \log f_{j, \nu } (x_{\nu } ) ,
\end{equation}  
and the classifier of Na\" ive Bayes, once calibrated or trained, is again defined by the MAP-rule (\ref{eqlda1}).   
\medskip   
   
There remains the question of how to choose the univariate densities $f_{j, \nu } (x_{\nu } ) . $ There are two basic methods:   
   
\begin{enumerate}\label{NBestimation}   
   
\item {\it Parametric specification}: one can simply specify a parametric family of univariate distributions for each of the $f_{j , \nu } $ and estimate the parameters from a training sample.    
Note that NB with normal $f_{j, \nu } $ reduces to the special case of QDA with diagonal variance-covariance matrices $V_j . $      
\medskip

\item {\it Non-parametric specification}: alternatively, one can employ a non-parametric estimator for the $f_{j , \nu } $ such as the Kernel Density Estimator (KDE).   
We recall that if $\{ x_1, x_2, \ldots , x_n \} $ is a sample of a real-valued random variable $X $ with probability density $f $, then {\it Parzen's Kernel Density Estimator} of $f $ is defined by   
\begin{equation}
\widehat{f } (x) = \frac{1}{n b } \sum_{i=1}^n K \left( \frac{x-x_i }{b } \right) ,   
\end{equation}
where the \textit{kernel} $K $ is a non-negative function on $\mathbb{R } $ whose integral is 1 and whose mean is 0, and $b > 0 $ is called the \textit{bandwidth parameter}.       
\end{enumerate}   
Kernel Density Estimators represent a way to smooth out sample data. The choice of $b $ is critical: too small a $b $ will lead to possibly wildly varying $\widehat{f } $ which try to follow the data too closely, while too large a $b $ will "oversmooth" and ignore the underlying structure present in the sample. Three popular kernels are the Normal kernel, where $K(x) $ is simply taken to be the standard normal pdf, and the so-called Triangular kernel and the Epanechnikov kernel (Epanechnikov, 1969), which are compactly supported piece-wise polynomials of degree 1 respectively 2, and for whose precise definition we refer to the literature. For simplicity we have used the same kernel and bandwidths for all $f_{j, \nu } . $   
   
Our Intra-classifier comparison results show that the KDE estimator with Normal kernel outperforms other the two kernel functions in most cases. Regarding the choice of kernels and of bandwidths, for us the issue is not so much whether the KDE estimator provides good approximations of the individual $f_{j , \nu } $'s 
but rather how this affects the classification error: in this respect, see Figure \ref{PerformanceNB}.   
\medskip

\subsubsection{An Illustrative Example for Na\"ive Bayes}
\begin{figure}[h]
\caption{Naive Bayes Class Conditional Independence vs Correlated Gaussian} \label{NBgraph2}
\centering
\includegraphics[scale=0.5]{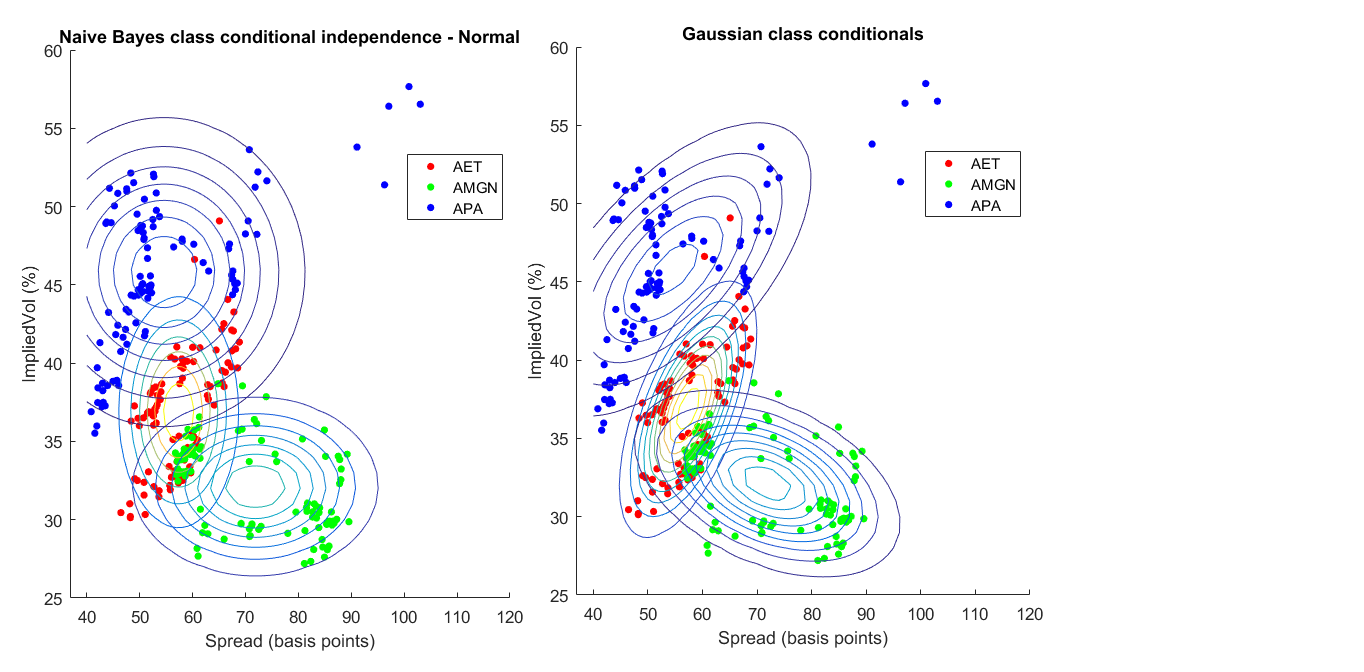} 
\end{figure}

Figure \ref{NBgraph2} compares the contour plots of the class conditional densities found by NB with normally distributed individual features (graph on the left)  with the ones found by QDA (graph on the right) for our example \ref{running_DA_example}.   
The three normal distributions on the left have principal axes parallel to the coordinate axes, reflecting the independence assumption of Na\'ive Bayes. They show a stronger overlap than the ones on the right, whose tilted orientation reflects their non-diagonal covariance matrices. The stronger overlap translates into higher misclassification rates, and our empirical study confirms that Na\"ive Bayes Classifiers perform less well than QDA for any of the feature variable selections used.    
   
As listed in Table \ref{classifiersfig}, we investigated the effects of Bandwith and Kernel function choices on Na\"ive Bayes' classification performance. Our empirical results in Section \ref{empsection} show that Na\"ive Bayes Classifiers perform rather poorly in comparison with the other Classifier families, in contrast with results from the literature on classification with non-financial data: see for example Rish {\it et al.} (2001) . This is probably due to the independence assumption of Na\"ive Bayes, which is not appropriate for financial data: as indicated in Figure \ref{PerfNBCorrelation}, for our data set, which came from a period of financial stress, roughly 80\% of the pairwise correlations of our feature variables are above 70\%. Even under normal circumstances one expects a 3-month historical volatility and a 3-month implied volatility to have significant dependence.   
The interested readers can find more details on the performance of NB in Appendix \ref{sectionindividaulclass}.   
   
\subsection{$k$-Nearest Neighbours}\label{sectionknn}
In Classification, there are typically two stages: the first is the Training stage where one estimates the parameters of the learning algorithm or, in Machine Learning parlance,  {\it trains} the algorithm. The second stage is called the Testing stage, where one uses the algorithm to classify feature vectors which were not in the Training set, and possibly checks the results against known outcomes, to validate the trained Classifier;  we will also speak of the Prediction or Classification stage.   
The \textit{$k$-Nearest Neighbours} or $k $-NN algorithm is an example of a so-called {\it lazy learning strategy}, because it makes literally zero efforts during the Training stage; as a result, it tends to be computationally expensive in the Testing/Prediction stage.

\subsubsection{The $k $-NN algorithm}   
   
Letting $D^T = \{ (\mathbf{x }_i , y_i ) : i = 1 , \ldots , n \} $ be, as before, our Training Set, where for us $\mathbf{x }_i $ is an observed feature vector of the observable counterparty $y_i $, the $k$-NN algorithm can be described as follows:   
\begin{itemize}   
   
\item For a given feature vector $\textbf{x } $, which we can think of as the feature vector of some non-observable name, compute all the distances $d(\textbf{x}, \textbf{x}_i ) $ for $(\mathbf{x }_i , y_i ) \in D^T $, where the metric $d $ can be any metric of one's choice on the feature space $\mathbb{R }^d $, such as the Euclidean metric or the so-called City Block metric.   
    
\item Rank order the distances and select the $k $ nearest neighbours of $\mathbf{x } $ amongst the $\mathbf{x }_i . $  Call this set of points $\Re (\mathbf{x } , k ) $: in Figure \ref{KNNexample} these are the points within the circle.   
   
\item Classify $\mathbf{x } $ to that element $\widehat{y }(\mathbf{x } ) \in \{ 1, 2, \ldots , N \} $ which occurs most often amongst the $y_i $ for which $\mathbf{x }_i \in \Re (\mathbf{x } , k ) $, with some arbitrary rule if there is more than on such an element (e.g. taking the smallest). This is called the \textit{Majority Vote Rule}; see Hastie {\it et al.} (2009) for a Bayesian justification.   
   
\end{itemize}   
   
As already mentioned, and in contrast in contrast to the other algorithms we considered,   
\medskip

\subsubsection{An Illustrative Example for $k$-NN }\label{KNNExample}   
   
\begin{figure}[h]
\caption{$k$--NN Illustrative Example}\label{KNNexample}
\centering
\includegraphics[scale=0.7]{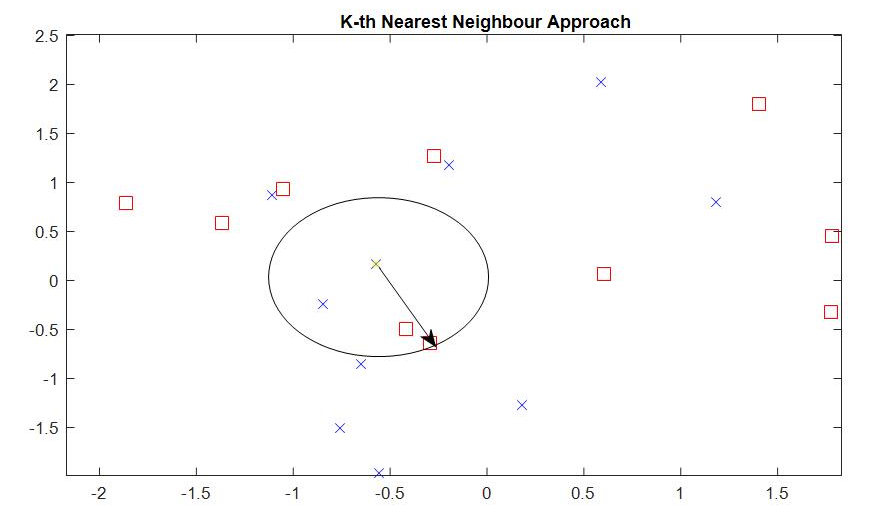} 
\end{figure}

Figure \ref{KNNexample} provides a simple illustration of the \textit{$k$}-th Nearest Neighbour algorithm. Based on a 2-dimensional feature space, it depicts the feature vectors of two observables, each with 10 data samples inside a rectangular box: one observable is represented by blue ''\textbf{x}'' shapes; the other by red ''\textbf{$\Box$}'' shapes. Suppose we want to use $k$-NN to classify the nonobservable $\mathbf{x } $ represented by the grey cross. %denoted by $x_q$ based on the data inside the rectangular box.   
Taking $k=3 $ and the Euclidean distance as metric, we find that the third nearest point to $\mathbf{x } $ within the Training set happens to be the red ''\textbf{$\Box$}'' to which the arrow points. Amongst these three nearest neighbours, the red boxes occur twice and the blue boxes only occur once. By  the Majority Vote Rule, the CDS-Proxy for $\mathbf{x } $ is then selected to be the (name represented by) the red boxes.       

As indicated in Table \ref{classifiersfig}, we investigated $k $-NN with three different distance metrics, the Euclidean metric, the City  Block or $\ell ^1 $-metric and the so-called Mahalanobis distance which takes into account the spatial distribution (spread and orientation) of the feature vectors in the training sample\footnote{The explicit formulas are: $d_2 (\mathbf{x } , \mathbf{y } ) = \sqrt{(\mathbf{x } - \mathbf{y } )^T (\mathbf{x } - \mathbf{y } } = \sqrt{\sum _{\nu } (x_{\nu } - y_{\nu } )^2 } $ for the Euclidean metric, $d_1 (\mathbf{x } , \mathbf{y } ) = \sum _{\nu } |x_{\nu } - y_{\nu } | $ for the City Block metric, and $d_V = \sqrt{ (\mathbf{x } - \mathbf{y } )^T \widehat{V }^{-1 } (\mathbf{x } - \mathbf{y } ) } $ for the Mahalanobis metric, where $\widehat{V } $ is the empirical variance-covariance matrix of the $\mathbf{x }_i $'s. }, and we studied the dependence of the classification accuracy on  $k . $ The Intra-classifier results are presented in figure \ref{knnperffigure} and in table \ref{KNNperftbl2} of Appendix \ref{sectionindividaulclass}, and the comparison of $k $-NN with the other Classifier families is done in Section \ref{empsection}.   
\subsection{Logistic Regression}\label{sectionLR}
In the Bayesian-type classifiers discussed so far, we have estimated the posterior probability densities $\mathbb{P } ( j \mid \mathbf{x } ) $ %$:= \mathbb{P } (Y = j \mid \textbf{X } = \mathbf{x } ) $   
by first modelling and estimating the  feature vectors' class-conditional probability distributions, together with the prior probabilities for class membership, and  then applying Bayes Formula. By contrast, Logistic Regression (LR) straightforwardly assume a specific functional form for $\mathbb{P } (j \mid \textbf{x } ) $ which then is directly estimated from the training data. Logistic Regression classifiers come in two types, Binomial and Multinomial, corresponding to two-class and multi-class classification.   
   
\subsubsection{Binomial LR Classification}   
   
Suppose that we have to classify a feature vector $\mathbf{x } \in \mathbb{R }^d $ to one of two classes, $Y = 0 $ or $1 . $ Binomial Logistic Regression assumes that the probability that $Y = 1 $ given $\mathbf{x } $ is given by 
\begin{equation}\label{eqlr}
p(\mathbf{x } ; \boldsymbol{\beta } ) := \mathbb{P } (Y = 1 \mid \textbf{x}) := g\left( \beta _0 + \sum _{\nu } \beta _{\nu } x_{\nu } \right) = 
g\left( \boldsymbol{\beta }^T \tilde{\textbf{x} } \right) ,   
%\frac{1}{1+e^{-\boldsymbol{\beta^T} \textbf{x}}} ,   
\end{equation}
where $g(z) =\left( 1 + e^{-z } \right)^{-1 } $ is the Logistic or Sigmoid function, $\boldsymbol{\beta}= ( \beta_0 , \beta_1, \beta_2, \dots,\beta_d ) $ is a vector of parameters and $\tilde{\mathbf{x } } := (1, \mathbf{x } ) $, the component 1 being added to include the intercept term $\beta _0 . $  Given a training set $D^T = \{ (\mathbf{x }_i , y_i ) \mid i = 1 , \ldots , n \} $ with $y_i \in \{ 0 , 1 \} $, the likelihood of obtaining the data of $D^T $ under this model can readily be written down, and $\boldsymbol{\beta } $ can be determined by Maximum likelihood Estimation (MLE), for which standard statistical packages are available.   
   
Once calibrated to parameters $\widehat{\boldsymbol{\beta } } $ we classify a new feature vector $\mathbf{x } $ to $Y = 1 $ if $p(\mathbf{x } , \widehat{\boldsymbol{\beta} } ) \geq 0.5 $, and to $Y = 0 $ otherwise.: equivalently, since both probabilities sum to 1,   
$$   
\widehat{y }(\mathbf{x } ) = \arg \max _j  p_j (\mathbf{x } , \widehat{\boldsymbol{\beta } } ) ,   
$$   
where $p_1 (\mathbf{x } , \widehat{\boldsymbol{\beta } } ) := p(\mathbf{x } , \widehat{\boldsymbol{\beta} } ) $ and $p_0 (\mathbf{x } , \widehat{\boldsymbol{\beta} } ) = 1 - p(\mathbf{x } , \widehat{\boldsymbol{\beta} } ) $ and where we agree to classify $\mathbf{x } $ to 1 if both probabilities are equal.   
   
\subsubsection{Multinomial LR Classification}   
 
To extend the two-class LR algorithm to the multi-class case, we recast a multi-class classification problem as a sequence of two-class problems. For example, we can single out one of the observables $j \in \{ 1 , \dots , N \} $ as a reference class, and successively run a sequence of Binomial Logistic Regressions between membership or no-membership of the reference class, that is, we take $Y = 1 $ if $\mathbf{x } $ belongs to this class $j $, and $Y = 0 $ otherwise. This will result in $N-1$ logistic regressions with $N-1 $ parameter vectors $\widehat{\boldsymbol{\beta } }_j $, $j = 1, \ldots , N -1. $ The likelihoods that a new feature variable $\mathbf{x } $ will be classified to $j $ then is $p (\mathbf{x } , \widehat{\boldsymbol{\beta } }_j ) $ and we classify $\mathbf{x } $ to that class for which this likelihood is maximal. In other terms, we define the classifier function $\widehat{y}(\mathbf{x}) $ by   
\begin{equation} \label{decisionMLR}
\widehat{y } (\mathbf{x } ) = \arg \max _j p (\mathbf{x } , \widehat{\boldsymbol{\beta } }_j ) ,   
\end{equation}
with, as usual, some more or less arbitrary rule for the exceptional cases when the maximum is attained for more than one value of $j $, such taking the largest or smallest such $j . $   
\medskip   
   
\noindent {\bf Remark}. In another version of multi-class Logistic Regression one directly models the conditional probabilities by   
\begin{equation} \label{eq:multi_logistic}   
\mathbb{P } (j | \mathbf{x } ) = \frac{e^{\boldsymbol{\beta }_j ^T \tilde{\mathbf{x } } } }{\sum _l e^{\boldsymbol{\beta }_l ^T \tilde{\mathbf{x } } } } \ , \ j = 1, \ldots , N,    
\end{equation}   
where $\tilde{\mathbf{x } } := (1, \mathbf{x } ) $ as before, and each $\boldsymbol{\beta } $ is a parameter vector as above.
The $N $ parameter vectors can be estimated by Maximum Likelihood and new feature vectors are classified to the class for which (\ref{eq:multi_logistic}) is maximal. If $N = 2 $, this model is equivalent to Binary Logistic Regression, with parameter vector $\boldsymbol{\beta } = \boldsymbol{\beta }_1 - \boldsymbol{\beta }_2 $, if we let $j = 1 $ correspond to $Y = 1 . $ More generally we can translate the $\boldsymbol{\beta }_i $'s by a common vector without changing the probabilities (\ref{eq:multi_logistic}) so we can always suppose that $\boldsymbol{\beta }_N = 0 . $ We have not used this particular version, but (\ref{eq:multi_logistic}) will re-appear below as the Output Layer of a Neural Network Classifier.

\subsection{Decision Trees} \label{sectionDT}   
   
The Decision Tree algorithm essentially seeks to identify rectangular regions in feature space which characterise the different classes (for us: the observable counterparties), to the extent 

that such a characterisation is possible. Here "rectangular" means that these regions are going to be defined by sets of inequalities $a_1 \leq x_1 < b_1, \ldots, x_d \leq x_d < b_d $, where $a_{\nu } $ and $b_{\nu } $ may be $- \infty $ respectively $\infty . $ These regions are found by a tree-type construction in which we successively split the range of values of each of the feature variables into two subintervals, for each subinterval determine the relative frequencies of each of the observable counterparties having its feature variable in the subinterval, and finally select that split of that component for which the separation of the observables into two subclasses becomes the "purest", in some suitable statistical sense whose intuitive meaning should be that the empirical distribution of the counterparties associated to the subintervals becomes more concentrated around a few single ones. This procedure is repeated until we have arrived at regions which only contain a single class, or until some pre-specified constraints on Tree Size, in terms of maximum number of splits, has been reached. The Decision Tree is an example of a ''greedy'' algorithm where we seek to achieve local optimal gains, instead of trying to achieve some global optimum.   
   
Historically, various types of tree-based algorithms have been proposed in Machine Learning. The version used in this paper is a binary decision tree similar to both the Classification and Regression Tree (CART), originally introduced by Breiman {\it et al.} (1984), and to the C4.5 proposed by Quinlan (1993). If needed, the tree can be pruned, by replacing nodes or removing subtrees while checking, using cross validation, that this does not reduce the predictive accuracy of the tree.         
   
\subsubsection{The Decision Tree algorithm}   
   
For the construction of the decision tree we need a criterion to decide which of two sub-samples of $D^T $ is more concentrated around a (particular set of) counterparties. This can be done using the concept of an {\it impurity measure} , which is a function $G $ defined on finite sequences of probabilities $\mathbf{p } = (p_1 , \ldots , p_N ) $, where $p_j  \geq 0 $ and $\sum _j p_j = 1 $, which has the property that $G(\mathbf{p } ) $ is minimal iff all $p_i $'s except one are 0, the remaining $p_i $ then necessarily being 1; one sometimes adds the conditions that $G $ be symmetric in its arguments and that $G $ assumes its maximum when all $p_k $'s are equal: $p_1 = \cdots = p_N = 1 /N . $ Two popular examples of impurity measure which we also used for  our study are:      
   
\begin{enumerate}
\item the \textit{Gini Index},      
\begin{equation}
G = 1 - \sum _{j = 1 }^N p_j ^2 ,    
\end{equation}   
which, by writing it as $\sum_{j = 1 }^N p_j (1 - p_j ) $, can be interpreted as is the sum of the variances of $N $ Bernoulli random variables whose respective probabilities of successes are $p_j $, and   
   
\item the \textit{Cross Entropy}, %defined as   
\begin{equation}
G=-\sum_{j = 1 }^N p_j \log p_j .   
\end{equation}   
   
\end{enumerate}   
Splitting can be done so as to maximize the gain in purity as measured by $G . $ Another somewhat different splitting criterion is that of {\it Twoing}, which will be explained below. The Decision Tree is then constructed according to the following algorithm %for a given training sample $D^T $:   
   
\begin{enumerate}
\item	Start with the complete training sample $D^T $ at the root node $T_0 . $   
   
\item Given a node $T_p $ (for "parent") with surviving %
sample set $D^{T _p }  $, for each couple $s = (\nu ,  r ) $ with $ 1 \leq \nu \leq d $ and  $r \in \mathbb{R } $, split $D^{T _p }  $ into two subsets,  the set $D^{T _p } _ L (s) $ of data points $(\mathbf{x }_i , y_i ) \in D^{T _p }  $ for which the $\nu $-th component $x_{i , \nu } < r $, and the set $D^{T _p } _R (s) $ defined by $x_{i , \nu } \geq r . $ We will call $s $ a {\it split}, and $D^{T _p } _ L (s) $ and $D^{T _p } _ R (s) $ the associated left and right split of $D^{T _ p } $, respectively. Observe that we can limit ourselves to a finite number of splits, since there are only finitely many feature values $x_{i , \nu } $ for $(\mathbf{x }_i , y_i ) $ in $D^{T _p }  $, and we can choose the $r $'s arbitrarily between two successive values of the $x_{i , \nu } $'s, for example half-way between.         
\bigskip

\item For $j = 1, \ldots , N $, let $\pi _{p , j } $ be the proportion of data points $(\mathbf{x }_i , y_i ) \in D^{T _p }  $ for which $y_i = j $, and, similarly, for a given  split $s $ let $\pi _{L , j } (s) $ and $\pi _{R , j } (s) $ be the proportion of such points in $D^{T _p } _L (s) $ and $D^{T _p } _R (s) . $ Collecting these numbers into three vectors  $\boldsymbol{\pi }_p (s) = \left( \pi _{p, 1 } (s) , \ldots , \pi _{p , N } (s) \right) $, $\boldsymbol{\pi }_L (s) = \left( \pi _{L , 1 } (s) , \ldots , \pi _{L , N } (s) \right) $ and similarly for $\boldsymbol{\pi }_R (s) $, compute each splits {\it purity gain}, defined as   
$$   
\Delta G (s) := G( \boldsymbol{\pi }_p ) - \left( p_L (s) \, G (\, \boldsymbol{\pi }_{p , L } (s) ) \, ) \, + \,  p_R (s) \, G (\, \boldsymbol{\pi }_{p , R } (s) \, )  \right) ,   
$$   
where $p_L (s) := \# D^{T _p } _L (s) / \# D^{T _p }  $ and $p_R (s) := \# D^{T _p } _R (s) / \# D^{T _p }  $ are the fractions of points of $D^{T _p }  $ in the left and right split of $D^{T _p } $, respectively.

\item Finally, choose a split $s^* $ for which the purity gain is maximal\footnote{$s^* $ is not necessarily unique though generically one expects it t be} and define two daughter nodes $T_{p, L } $ and $T_{p, R } $ with data sets $D^{T _p }_L (s^* ) $ and $D^{T _p } _R (s^* ) . $   

\item	Repeat steps 2 to 4 until  each new node has an associated data set which only contains feature data belonging to a single name $j $, or until some artificial stopping criterion on the number of nodes is reached.   
   
\end{enumerate}   
It is clear that the nodes can in fact be identified with the associated data sets. If we use {\it twoing}, then step 3 is replaced by computing   
$$  
p_L (s ) p_R (s) \, \left( \sum _{j = 1 }^n \, \left | \, \pi _{j , R } (s) - \pi _{j , L } (s) \, \right | \, \right)^2 ,   
$$   
and step 4 by choosing a split which maximizes this expression.   
\medskip   
   
One advantage of tree-based methods is their intuitive content and easy interpretability. We refer to the number of leafs in the resultant tree as the \textit{tree size} or the \textit{complexity of the tree}. Oversized trees become less easy to interpret. To avoid such overly complex trees, we can prescribe a bound on the number of splits $z $ as a stopping criterion. We can search for the optimal choice of tree size by examining the cross-validation results across a range of maximum splits. As shown in the section on empirical results, the classification accuracy is not strongly affected by $z $ anymore once it has reached the level of about 20.

\subsubsection{An Example of a Decision Tree}
Table \ref{DTfig1} shows the decision rules generated by the Decision Tree algorithm using as feature vector $\mathbf{x } := (PD_{3yr}, PD_{5yr}, \sigma ^{\rm h } _{3m } ) $, for the five observable counterparties indicated by their codes in the last column of the table.   

The algorithm was run on data collected from the 100 days leading up to Lehman's bankruptcy on 15-Sept-2008: see Appendix \ref{sixmodels}.   
   
The tree
has nine nodes, labelled from 1 to 9. Depending on the values of its feature variables, a nonobservable will be led through a sequence of nodes starting from node 1, until ending up with a node associated with a single observable counterparty, 
to which it then is classified.   
\smallskip   
   
\begin{table}[h]
\caption{A Simple Illustrative Example of Decision Tree based CDS Proxy Method}\label{DTfig1}
\centering    
\includegraphics[scale=0.65]{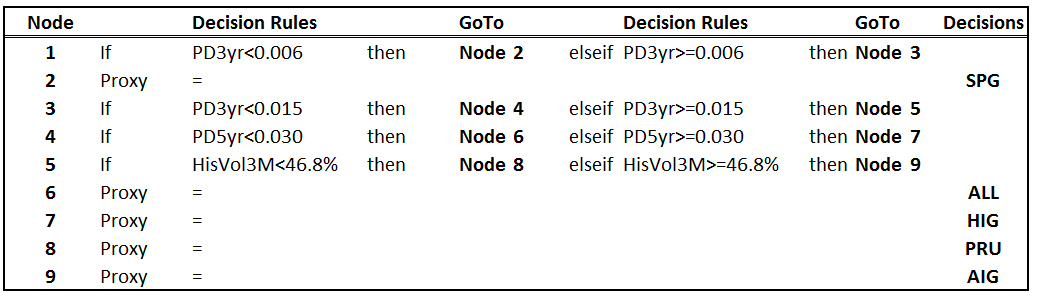} \\
\end{table}

As shown in Table \ref{classifiersfig}, we have investigated the impact of tree-size and of the different definitions of purity gain (Gini, Entropy, Twoing) on the Decision Tree's classification performance:  
see section \ref{empsection} for the Cross-classifier comparison, and Figure \ref{dtgraph1} and its associated table for the Intra-classifier comparison.   
   
It is known that the Decision Tree algorithm may suffer from overfitting: it may perform well on a training set but fail to achieve satisfactory results on test sets. For that reason we have also investigated the so-called \textit{Bootstrapped Aggregated Trees} or \textit{Bagged Trees}, an example of an Ensemble Classifier which we discuss in further detail in Section \ref{sectionbaggedtree} below.   

\subsection{Support Vector Machines}\label{sectionSVM}%
We will limit ourselves to an intuitive geometric description of the Support Vector Machine (SVM) algorithm, referring to the literature for technical details: see for example Hastie {\it et al.} (2009).   
   
Traditionally, the explanation of a SVM starts with the case of a two-class classification problem, with classes $y = 1 $ and $y = - 1 $, for which the feature vector components of the training data $D^T = \{ (\mathbf{x }_i , y_i ) \in \mathbb{R }^d \times \{ \pm 1 \} , \, i = 1 , \ldots , n \, \} $ can be {\it linearly separated} in the sense that we can find a hyperplane $H \subset \mathbb{R }^n $ such that all date $\mathbf{x }_i $ for which $y_i = 1 $ lie on one side of the hyperplane, and those for which $y_i = - 1 $ lie on the other side. Needless to say, for a given data set, the assumption of linear separability is not necessarily satisfied, and the case when it isn't will be addressed below. If it does hold, one also speaks of the existence of a {\it hard margin}.   
   
Assuming such a hard margin, the idea of a SVM is to choose a separating hyperplane which maximises the distances to both sets of feature variables, those for which $y_i = 1 $ and those for which $y_i = - 1 . $ The two distances can be  made equal, and their sum $M $ is called the {\it margin}: see Figure \ref{svmgraph1}.
Using some elementary analytic geometry, this can be reformulated as a quadratic optimisation problem with linear inequality constraints:   
\begin{equation} \label{eq:SVM_1}   
\left \{ \begin{array}{ccc}   
\min _{\boldsymbol{\beta } , \beta_0 } || \boldsymbol{\beta } ||^2 \\   
\mbox{subject to } \\   
y_i \left( \boldsymbol{\beta }^T \mathbf{x}_i + \beta _0 \right) \geq 1 , \ i = 1 , \ldots , n .   
\end{array}   
\right.   
\end{equation}

Data points for which the inequality constraint is an equality for the optimal solution are called {\it support vectors}: these are the vectors which determine the optimal margin. If $(\boldsymbol{\beta }^* , \beta _0 ^* ) $ is the, unique, optimal solution, the any new feature-vector $\mathbf{x } $ is assigned to the class $y = 1 $ or $y = - 1 $   according to whether $\widehat{y } (x) $ is positive or negative, where       
\begin{equation}   
\widehat{y } (x) = \boldsymbol{\beta }^{* T } \mathbf{x } + \beta _0 ^* .   
\end{equation}   
The bigger $| \widehat{y } (x) | $ the more "secure" the assignation of the new data point $\mathbf{x } $ to its respective class, a point to keep in mind for the extension of the algorithm to multi-class classification below.   

\subsubsection{An illustration of Margin}\label{svmIllustration}
\begin{figure}[h]
\caption{SVM Illustrative Example for Margin}\label{svmgraph1}
\centering
\includegraphics[scale=0.5]{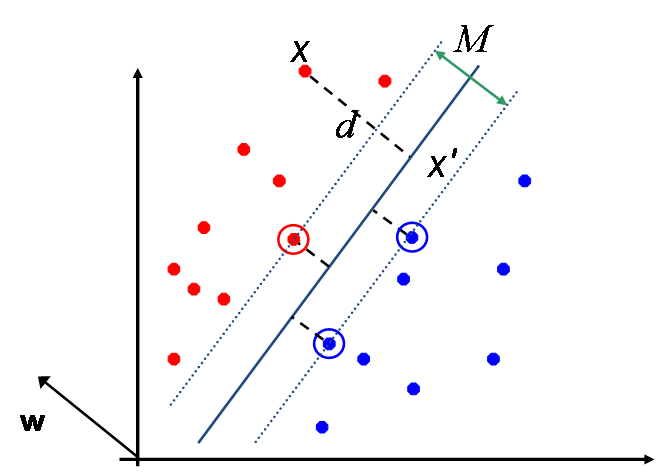}
\end{figure}

Figure \ref{svmgraph1} illustrates the concept of linearly separable data with maximal margin $M . $ 
  
\subsubsection{Non-linearly separable data}   
   
If the feature data belonging to the two classes are not linearly separable, they can always be separated by some curved hyper-surface $S $, and the data become linearly separable in new coordinates $(\xi _1 , \ldots , \xi _d ) $ in which for example the equation of $S $ reduces to $\xi _1 = {\rm constant } . $ A standard example is that of a set of points in the interior of a sphere of radius $R $ versus another set of points outside of the sphere: these are evidently not linearly separable in the usual Cartesian coordinates, but will become separable in polar coordinates.   
%These cannot be linearly separated, but by going over to spherical coordinates they become linearly separable: if $r = ||x || $, they are separated by the hyperplane $\{r = R \} . $   
More generally, one can always find an invertible smooth map $\varphi $ from $\mathbb{R }^d $ into some $\mathbb{R }^k $ with $k \geq d $ such that the transformed feature-vectors $\varphi (\mathbf{x }_i ) $ become linearly separable. One can then run the algorithm in $\mathbb{R }^k $ on the transformed data set $\{ (\varphi (\mathbf{x }_i ) , y_i ) : i =1 , \ldots , N \} $, and construct a decision function of the form $\widehat{y }(x) = \boldsymbol{\beta }^{* T } \varphi (\mathbf{x } ) + \beta _0 ^* $ which can be used for classification.      
   
From a theoretical point of view this is very satisfactory, but from a practical point there remains the problem on how to let the machine automatically choose an appropriate  map $\varphi . $ To circumvent this we first consider the dual formulation of the original margin maximisation problem (\ref{eq:SVM_1}). It is not difficult to see that the optimal solution can be written as a linear combination $\boldsymbol{\beta }^* = \sum _{i = 1 } ^n \alpha _i ^* \mathbf{x }_i $ of the data points: any non-zero component $\boldsymbol{\beta }_{\perp } $ of $\boldsymbol{\beta } $ perpendicular to the $\mathbf{x }_i $ would play no r\^ole for the constraints, but contribute a positive amount of $|| \boldsymbol{\beta}_{\perp } ||^2 $ to the objective function. In geometric terms, if all data lie in some lower-dimensional linear subspace (for example a hyperplane), then the optimal, margin-maximising, separating hyperplane will be perpendicular to this sub-space. One can therefore limit oneself to $\boldsymbol{\beta } $'s of the form $\boldsymbol{\beta } = \sum _i \alpha _i \mathbf{x }_i $, and instead of (\ref{eq:SVM_1}) solve   
\begin{equation} \label{eq:SVM_3}      
\left \{ \begin{array}{ccc}   
\min _{\boldsymbol{\alpha } ,  \alpha_0 } \sum _{i , j } \alpha _i \alpha _j \mathbf{x }_j ^T \mathbf{x }_i \\     
\mbox{s. t. } \\   
y_i \left( \sum _j \alpha _j \mathbf{x } _j ^T \mathbf{x }_i + \alpha _0 \right) \geq 1 , \ i = 1 , \ldots , n .   
\end{array}   
\right.   
\end{equation}   
For the transformed problem we simply replace the inner products $\mathbf{x }_j ^T \mathbf{x }_i $ by $\varphi (\mathbf{x _j } )^T \varphi (\mathbf{x }_i ) $: note that the resulting quadratic minimisation problem is always $n $-dimensional, irrespective of the dimension $k $ of the target space of $\varphi . $ Now the key observation is that the symmetric $n \times n $-matrix with coefficients   
\begin{equation} \label{eq:Mercer}   
k(\mathbf{x }_i , \mathbf{x }_j ) = \varphi (\mathbf{x }_j )^T \varphi (\mathbf{x }_i )   
\end{equation}   
is positive definite\footnote{meaning that for all vectors $(v_1 , \ldots , v_N ) $, $\sum _{i , j } k(\mathbf{x }_i , \mathbf{x }_j ) v_i v_j \geq 0 $ }, and that, conversely, if $k (\mathbf{x } , \mathbf{y } ) $ is any function %on $\mathbb{R }^n \times \mathbb{R }^n $, the Cartesian product of feature space with itself   
for which the matrix $\left( k( \mathbf{x }_i , \mathbf{x }_j )\right) _{i, j } $ is positive definite, then it can be written as (\ref{eq:Mercer}) for an appropriate $\varphi $, by a general result known as Mercer's theorem. Functions $k $ for which $\left( k( \mathbf{x }_i , \mathbf{x }_j )\right) _{i, j } $ is always positive definite, whatever the choice of points $\mathbf{x }_i $, are called positive definite kernels. Examples of such kernels include the (non-normalised) Gaussians $k(\mathbf{x } , \mathbf{y } ) = e^{ - c || \mathbf{x } - \mathbf {y } ||^2 } $ and the polynomial kernels $\left( 1 + \mathbf{y }^T \mathbf{x } \right)^p $, where $p $ is a positive integer.      
   
To construct a general non-linear SVM classifier, we choose a positive definite kernel $k $, and solve   
\begin{equation} \label{eq:SVM_4}   
\left \{ \begin{array}{ccc}   
\min _{\boldsymbol{\alpha } ,  \alpha_0 } \sum _{i , j } \alpha _i \alpha _j \, k \left( \mathbf{x }_i , \mathbf{x }_j \right) \\   
\mbox{s. t. } \\   
y_i \left( \sum _j \alpha _j k \left( \mathbf{x }_i , \mathbf{x }_j \right) + \alpha _0 \right) \geq 1 , \ i = 1 , \ldots , n .   
\end{array}   
\right.   
\end{equation}   
The trained classifier function then is the sign of $\widehat{y }(\mathbf{x } ) $, where   
$$   
\widehat{y } (\mathbf{x } ) := \sum _{j = 1 } ^n \alpha _j ^* k \, \left( \mathbf{x } , \mathbf{x }_j \right) + \alpha _0 ^* ,   
$$  
the $^* $ indicating the optimal solution.   
   
\subsubsection{Hard versus soft margin maximisation} Although linear separation is always possible after transformation of coordinates, it may be advantageous to allow some of the data points to sit on the wrong side of the separating surface, if we do not want the latter to behave too "wildly": think of the example of two classes of points, "squares" and "circles", with all the "circles" at distance larger than 1 from 0, and all the "squares" at distance less than 1, except for one, which is at distance 100.   
   
Also, even if the data can be linearly separated, it may still be advantageous to let some of the data to be miss-classified, if this allows us to increase the margin, and thereby better classify future data points. We therefore might want to allow some miss-classification, but at a certain cost. This can be implemented by replacing the 1 in the right hand side in the $i $-th inequality constraints by $1 - \xi _i $, adding a cost function $ C \sum _i \xi _i $ to the objective function which is to be minimised, and minimising also over all $\xi _i \geq 0 . $

\subsubsection{Multiclass classification} We have given the description of the SVM classifier for two classes, but we still have to explain how to deal with a multi-class classification problem where we have to classify a feature vector $\mathbf{x } $ among $N $ classes.  %$\{ 1, 2, \ldots , N   
There are two standard approaches to this: we can break up the problem into $N $ two-class problems by classifying a feature vector as belonging to a given class or not belonging to it, for each of the classes. The two-class algorithm then provides us then with $N $ classifiers functions $\widehat{y }_j (\mathbf{x } ) $, $j = 1 , \ldots , N $,  which we then use to construct a global classifier by taking the (or a) $j $ for which $\widehat{y }_j (\mathbf{x } ) $ has maximum value (maximum margin). The other approach is to construct SVM classifiers for each of the $N (N - 1)/2 $ pairs of classes and again look select the one for which the two-class decision function has maximal value.   
\medskip   
   
As indicated in Table \ref{classifiersfig}, we investigated the SVM algorithm with Linear, Gaussian and Polynomial kernel and tested their performance for our CDS-Proxy problem. The results are presented in Section \ref{empsection} and Appendix \ref{sectionindividaulclass}.   
   
\subsection{Neural Networks}\label{sectionNN}
\subsubsection{Description}
Motivated by certain biological models of the functioning of a human brain and of its constituting neurons, Neural Networks represent a learning process by a network of stylised (mathematical models of) single neurons, organised into an Input Layer, and Output Layer and one or more intermediate Hidden Layers. Each single "neuron" transforms an input vector $\mathbf{z } = (z_1 , \ldots , z_p ) $ into a single output $u $ by first taking a linear combination $\sum_i w_i z_i $ of the inputs, %of the components $z_i $ of the input with weights $w_i $,   
adding a constant or bias term $w_0 $, and finally applying a non-linear transformation $f $ to the result:   
\begin{equation}  
u = f \left( \sum w_i z_i + w_0 \right) = f \left( \mathbf{w }^T \mathbf{z } + w_0 \right) ,   
\end{equation}   
where the weights $w_i $ of all of the neurones will be "learned" through some global optimisation procedure.   

The original idea, for the so-called  {\it perceptron}, was to take for $f $ a threshold function: $f(x) = 1$ if $x \geq a $ and 0 otherwise, which would only transmit a signal if the affine combination of input signals $\mathbf{w }^T \mathbf{z } + w_0 $ was sufficiently strong. Nowadays one typically takes for $f $ a smooth differentiable function such as the sigmoid function $\sigma $ defined by     
\begin{equation} 
\sigma (x) = \frac{1 }{1 + e^{- c x } }   
\end{equation}   
with $c $ an adaptable parameter. Other choices for $f $ are the hyperbolic tangent function or the linear function; these are all one-to-one, so nothing of the input signal is lost, contrary to the perceptron.   
   
As inputs of the neurons in the Input Layer one takes the feature vectors $\mathbf{x } . $ The outputs of the Input Layer neurons then serve as inputs for the neurons of the first Hidden Layer, whose outputs subsequently serve as inputs for the next Hidden Layer, etc.. Which output serves as input for which neuron depends on the network architecture: one can for example connect each of the neurons in a given Layer to all of the neurons in the next Layer. The outputs $\mathbf{u } ^f = (u_{\nu } ^f )_{\nu } $ of the final Hidden Layer undergo a final affine transformation to give $K $ values   
\begin{equation}   
\mathbf{w }_k ^T \mathbf{u }^f + w_{k0 } , \ \ k = 1 , \ldots , K .   
\end{equation}    
for certain weight vectors $\mathbf{w }_k = (w_{k \nu })_{\nu } $ and bias terms $w_{k0 } $ which, similar to the weights of the Hidden Layers, will have to be learned from test data: more on this below. For a regression with a Neural Network these would be the final output, but for a classification problem we perform a further transformation by defining   
\begin{equation}
\pi _k = \frac{e^{\mathbf{w }_k ^T \mathbf{u }^f + w_{k 0 } } }{\sum _{l = 1 } ^K e^{\mathbf{w }_l ^T \mathbf{u }^f + w_{l 0 } } } ,   
\end{equation}   
The interpretation is that $\pi _k $, which is a function of the input $\mathbf{x } $ as well as of the vector $\mathbb{W } $ of all the initial, intermediary, and final network weights, is the probability that the feature vector $\mathbf{x } $ belongs to class $k . $   
   
To train the network we note that minus the log-likelihood that an input $\mathbf{x }_i $ belongs to the (observed) class $y_i \in \{1 , \ldots , K \} $ is   
\begin{equation}   
- \sum _{i = 1 } ^N \sum _{k = 1 } ^K  \delta _{y_i , k } \log \pi _k (\mathbf{x }_i ; \mathbb{W } ) ,     
\end{equation}   
where $\mathbb{W } $ is the vector of all the weights and biases of the network. This is also called the cross-entropy. The weights are then determined so as to minimize this cross-entropy. This minimum is numerically approximated using a gradient descent algorithm. The partial derivatives of the objective function which are needed for this can be computed by backward recursion using the chain rule: this is called the backpropagation algorithm: see for example Hastie {\it et al.} (2009) for further details.   
   
The final decision rule, after training the Network, then is to assign a feature vector $\mathbf{x } $ to that class $k $ for which $\pi _k (\mathbf{x } , \widehat{\mathbb{W } } ) $ is maximal, where the hat indicates the optimised weights.    
   
\subsubsection{An Illustrative Example of A Simple Neural Network}
\begin{figure}[h]
\caption{An Illustration for A Simple Neural Network}\label{nngraph1}
\centering
\includegraphics[scale=0.8]{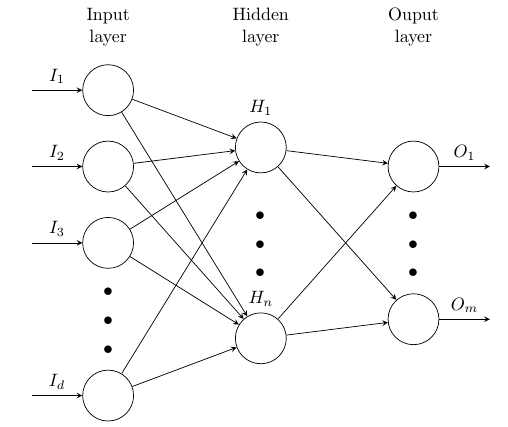}
\end{figure}

\textbf{Figure \ref{nngraph1}} shows a simple 3-Layer Neural Network including Input Layer ($d$ for \# of features), one Hidden Layer ($n$ for \# of Hidden Units) and Output Layer.   
     
\subsubsection{Parameterization}
\begin{figure}[h]
\caption{Activation Functions for Neural Network }\label{nnpara}
\centering
\includegraphics[scale=0.60]{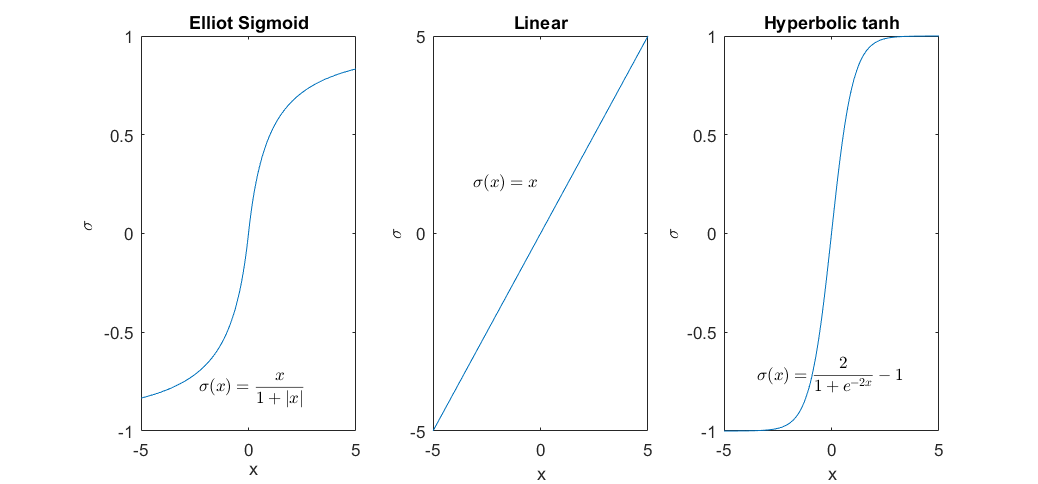}
\end{figure}

We restricted ourselves to Neural Networks with a single Hidden Layer, motivated by the {\it universal approximation theorem} of Cybenko and Hornik, which states that such networks are sufficient to uniformly approximate {\it any} continuous function on compact subsets of $\mathbb{R }^n . $  The remaining relevant parameters are then the activation function $f $ and the number of Hidden Units. As activation function we selected and compared the Elliot-sigmoid, purely linear and hyperbolic tangent functions: see Figure \ref{nnpara}. We also investigated the impact of the number of Hidden Units on Classification performance: the greater the number of these, the more complex the Neural Network, and the better, one might na\"ively expect, the performance should be. However, we found that, depending on Feature Selection, the performances for our proxy problem quickly stabilise for a small number of hidden neurons: see Figure \ref{NNperformfig}. We found the Neural Networks to be our best performing classifiers: see Section \ref{empsection} for further discussion.

\subsection{Ensemble Learning: Bagged Decision Trees} \label{sectionbaggedtree}
    
Bootstrapped Aggregating or {\it Bagging}, introduced by Breiman (1996), is based on the well-known bootstrap technique of Non-parametric Statistics (Efron 1979). Starting from a training set $D^T $ one generates new training sets $D_1 , \ldots , D_B $ by uniform sampling with replacement, and uses these to train classifiers $\widehat{y }_1 (\mathbf{x } ), \ldots , \widehat{y }_B (\mathbf{x } ) . $ The final classification is then done by {\it majority vote } (or {\it  decision by committee}): a feature vector $\mathbf{x } $ is associated to the class which occurs most often amongst the $\widehat{y }_i (\mathbf{x } ) . $ We will call $B $ the number of {\it learning cycles} of the bagging procedure.   

Breiman (1996) found that bagging reduces variance and bias. Similarly, Friedman and Hall (2000) report that bagging reduces variance for non-linear estimators such as Decision Trees. Bagging can be done using the same classifier algorithm at each stage, but can also be used to combine the predictions of classifiers from different classifier families. In this paper we have limited ourselves to bagging Decision Trees, to address the strong dependence of the latter on the training set and its resulting sensitivity to noise.

\subsubsection{An Example of Bagged Tree performance} 
\begin{figure}
\caption{Bagged Trees vs A Decision Tree}\label{BaggedTreegraph}   
\centering
\includegraphics[scale=0.9]{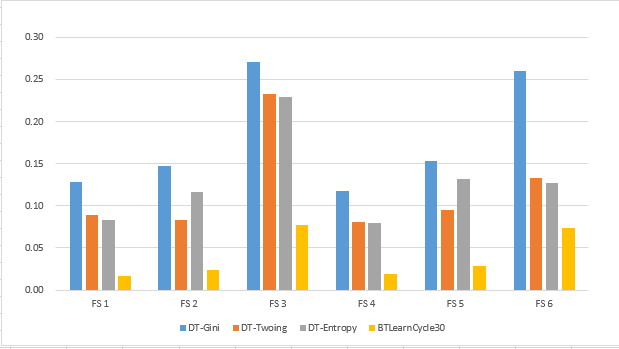}
\end{figure}

Figure \ref{BaggedTreegraph} show the improvement in performance, in terms of misclassification rates, from using the Bagged Tree as compared to the ordinary Decision Tree, for all three types of impurity measures (Gini, Twoing and Entropy). For this graph, the number of Learning Cycles $B $ was set to 30. We also investigated the dependence of the accuracy on $B $ and found that it stabilizes around $B = 30 $ for each of the Feature Selections: see Figure \ref{baggedtreeperffig} and Section \ref{section:Intra_Classifier_comparison} for further discussion. After bagging, the Decision Tree algorithm rose from sixth to third best performing classifier family: cf. Section \ref{CrossClassifierPerformance} below.    

\subsection{Statistical Procedures for Classifier Performance}\label{SectionKFold}
To examine performance of the various classifiers, we used the well established {\it $K$-fold Cross Validation} procedure, which is widely used in Statistics and Machine Learning.

\subsubsection{$K $-fold Cross Validation} Let $D^O $ be a set of observed data, consisting of feature vectors and the classes they belong to (for us: the observable counterparties).   

\begin{enumerate}   
   
\item  Randomise\footnote{to reduce sampling bias if the data come in a certain format, for example PD data ordering in increasing magnitude}   
$D^O$ and split it as a union of $K $ disjoint subsets  $D_n (K) $:  %where $n=1, 2, \ldots , K  $:   
$$   
D^O = \bigcup _{n=1 } ^K \, D_n (K) .     
$$   
Typically, the $D_n (K) $ will be of equal size. For \textit{Stratified $K$-fold Cross Validation}, each $D_n (K) $ is constructed in such a way that its composition, in terms of the relative numbers of class-specific samples which it contains, is similar to that of $D^O . $ %it is referred to as a Stratum.   
Stratified Cross Validation serves to limit sample bias in the next step.   
   
\item For $n=1,2,\dots,K$, define the {\it holdout sample} by $D^{H_n } = D_n (K) $ and  train the classifier on the $n $-th Training Set defined by   
\begin{equation}
D^{T_n } = D^O - D^{H_n } .   
\end{equation}   
Let $\widehat{y }_n $ denote the resulting classifier.      
%Where $D^{T_n}$ and $D^{H_n}$ are used for Training Set and Test Set for iteration $n$ respectively.   
   
\item For each $n $, test $\widehat{y }_n $ on the holdout sample $D^{H_n } $ by calculating the \textit{Misclassification Rate} $\epsilon_n^H $ defined by   \begin{equation}\label{misclassrate}
\epsilon_n^H=\frac{1}{\# D^{H_n}} \underset{(\mathbf{x } ,y)\in D^{H_n}  } {\sum } \, \left( \, 1 - I(y, \hat{y}(\mathbf{x } ) ) \, \right) ,   
\end{equation}   
where $I(u, v ) = 1 $ if $u = v $ and 0 otherwise.   
   
\item Take the sample mean and standard deviation of the $\epsilon _n ^H $ as empirical estimates for the {\it Expected Misclassification Rate} and its standard deviation by :   
\begin{align}\label{kfoldstats}
\widehat{u}_K=\frac{1}{K}\sum_{n=1}^K \epsilon_n^H \nonumber \\
\widehat{S}_K= \sqrt{\frac{1}{K} \sum_{n=1}^K (\epsilon_n^H-\widehat{u_K})^2} .   
\end{align}   
   
\end{enumerate}   
If we assume a distribution for the sampling error, such as a 
normal, a Student $t$ distribution, or even a Beta distribution (seeing that the $\varepsilon^H_n $ are all by construction between 0 and 1) we can translate  these numbers into a 95\% confidence interval, but we have limited ourselves to simply reporting 
$\widehat{u}_K$ and $\widehat{S}_K$. Also note that $1-\widehat{u }_K $ will be an estimate for the \textit{Expected Accuracy Rate}.

\subsubsection{Choice of \textit{K} for \textit{K}-fold cross validation}\label{kfoldcriteria} 
Kohavi (1995) recommends using Stratified Cross Validation to test Classifiers. Based on extensive datasets, it suggests that $K=10 $ is a good choice. This was also found by Breiman {\it et al.} (1984), who report that $ K=10$ gives satisfactory results for cross validation in their Decision Tree study. 
We examined the influence of $K $ on Stratified Coss Validation for the Discriminant Analysis, the Logistic Regression and the Support Vector Machine families (See Figures \ref{tblDAKFold}), \ref{LRperformfig} and \ref{SVMKFold}) and also found $K=10 $ to be a satisfactory choice, and unless otherwise stated, all of our cross validation results for the eight classifier families are obtained with $K=10 . $   

\subsection{Feature Selection and Feature Extraction}
After this discussion of the eight classifier families and of the statistical valuation procedure we use for assessing classifier performance, we turn to the feature variables. We discuss Feature Selection and Feature Extraction using PCA, and present an application of the latter.

\subsubsection{Feature Selection}      
Feature Selection can be based on purely statistical procedures such as the Forward, Backward and Stepwise Selections described in Hastie {\it et al.} (2009), can be done on theoretical grounds or can be informed by practice. In our study, we have taken the latter two routes, basing our selection of feature variables on own experience and on research such as that of Berndt {\it et al.} (2005), which reported that both probabilities of default (PD) and implied volatilities backed out from liquid equity option premiums of corporates have significant explanatory power for the CDS rates of corporates. For the PD data, Berndt {\it et al.} (2005)  used Moody's KMV\texttrademark \ Expected Default Frequency or EDF\texttrademark, which is obtained from Merton's classical Firm Value Model (Merton, 1974). In our study, we replaced these Expected Default Frequencies (which are only available to subscribers) by PD data from Bloomberg\texttrademark, which covers both public (Bloomberg, 2015) 
   
\subsubsection{Feature Extraction}   
   
Financial variables are typically strongly correlated, especially if they posses a term-structure, but also cross-sectionally, such as historical and implied volatilities of similar maturities. For our data set this is illustrated by the histogram of Figure \ref{PerfNBCorrelation} which shows the empirical distribution of pairwise correlations across the 16 feature variables and clearly indicates the very strong presence of significant correlations. It is well known that, for example, correlation amongst explanatory variables can have a strong impact on estimates of regression coefficients. The latter can undergo large variations if we perturb the model, by adding or deleting explanatory variables, or the data, by adding noise, and lead to erroneous estimates. This is known in Statistics as {\it Multicollinearity in Regression} and has been well researched: see for example Greene (1997). Performing a preliminary Principal Component Analysis (PCA) of the data and running the Regression in PCA space with only the first few of the principal Components can provide a solution to this problem. Mathematically, a PCA amounts to performing an orthogonal transformation on the feature vectors which diagonalises the variance-covariance matrix. The components of the transformed variables, which are referred to as the Principal Components or PCs, will then be uncorrelated; they are usually listed in decreasing order of their associated eigenvalues.   

In Machine Learning, the preprocessing of the original Feature Space by techniques such as PCA is referred to as \textit{Feature Extraction}; cf.  Hastie {\it et al.} (2009) . In fact, in areas such as image-recognition it is common practice to perform such a Feature Extraction procedure before proceeding to classification, to reduce the extremely large dimensionality of the feature space. For us, the dimension of feature space, being at most 16, is not an issue, but the presence of strong correlations between individual feature variables and its impact on classification might be. 
We have examined the impact of correlations by replacing the original feature variables by performing a PCA and using the PCs as input for the classification algorithms.  If correlation would strongly influence classification, then the classification results after PCA should be different, since the PCA components are, by construction, non-correlated. As we will see in Section \ref{CrossClassifierPerformance} below, for most classifiers families correlation doesn't influence classification, and where it does there are structural reasons.   
   
PCA is of course already often used in Finance, notably in Fixed Income and in Risk Management, mostly for dimension reduction purposes: see for example Rebonato (1999) as a general reference, and Brummelhuis {\it et al.} (2002) for an application to non-linear Value-at-Risk. In this paper we rather use it as a diagnostic tool, to ascertain the potential influence of feature correlations on classification.   
    
\subsubsection{An Example: Na\"ive Bayes Classification with and without PCA} \label{PCAstudyNB}   
   
\begin{figure}
\caption{Expected Accuracy Rates for Na\"ive Bayes Classifiers under PCA vs FS1 and FS6}\label{NaiveBayesPCAFigure}
\centering
\includegraphics[scale=0.75]{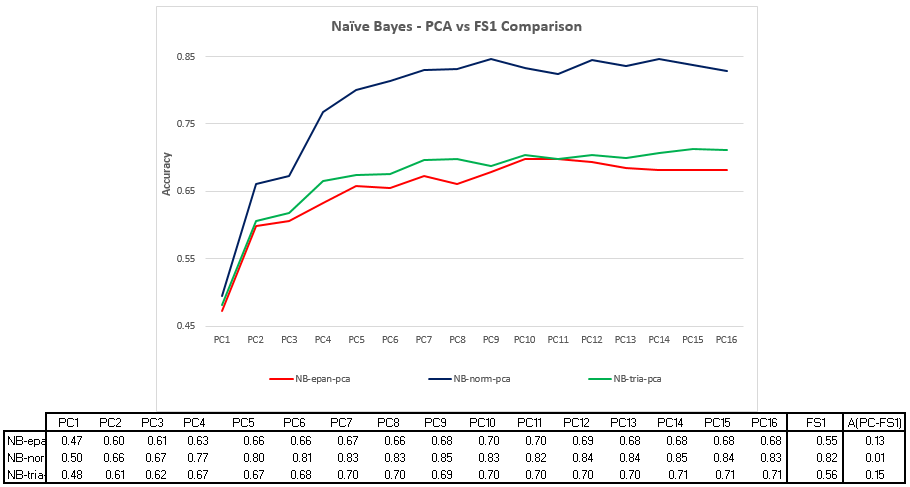}
\end{figure}

Although in applications the class-independence assumption made by Na\"ive Bayes classifiers is often violated, its performance with non-financial market data was cited as remarkably successful in Rish {\it et al.} (2001) . We compared the performance of Na\"ive Bayes with the original feature variables with that of Na\"ive Bayes using Principal Components, for the two feature vectors FS1 and FS. The graph in Figure \ref{NaiveBayesPCAFigure} plots the Empirical Accuracy Rates (as computed by $K $-fold Cross Validation) as a function of the number of PCs of FS1 which were used for the classification, while the first table lists the numerical values of these rates, as well as, in the final column, the Accuracy Rates obtained by using the "raw", non-transformed, FS1 variables. The second table presents the similar comparison results for Na\"ive Bayes with feature vector FS6.   

First of all, unsurprisingly, Figure \ref{NaiveBayesPCAFigure} shows that the greater the number of PCs used, the better the classification accuracy. Furthermore, classification on the basis of the full set of PC variables achieves a better accuracy than when using the non-transformed variables. This indicates that Na\"ive Bayes suffers from Multicollinearity in Regression issues, and also shows that PCA can be a useful diagnostic tool to uncover these. The explanation for the better accuracy after PCA can be found in the fact that our strongly correlated financial features fail to satisfy the independence assumption underlying NB, while the PCs meet this assumption at least approximatively, to the extent that they are at least uncorrelated.     
   
We also note that, looking at the graph, it takes between 7 and 10 principal components to achieve maximum accuracy, which is much more than the number of principal components, 1 or 2, needed to explain 99\% of the variance.  We can conclude from this that {\it variance explained is a poor indicator of classification performance}, a point which will be further discussed in Section \ref{CrossClassifierPerformance}.

\section{Empirical Performance Comparison Summmary}\label{empsection}   
   
In this section, we summarize our results from two angles:
\begin{itemize}
\item \textbf{Cross-classifier performance}, where we compare the classification performances of the eight classifier families with the six different feature selections listed in Appendix \ref{sixmodels}, individually and collectively, with and without feature extraction by PCA.   
      
\item \textbf{Intra-classifier performance}, in which for each of the eight classifier families individually we compare the individual performances of the different classifiers for different parameterisation choices and for the different feature selections, and discuss how the parameters for the cross-classifier comparison in section \ref{CrossClassifierPerformance} were set.   
   
\end{itemize}   
The different graphs and tables on which this summary is based are collected in Appendix \ref{sectionindividaulclass}.   

\subsection{Cross-classifier Performance Comparison Results}\label{CrossClassifierPerformance} 
\begin{figure}
\caption{Summary Classifier Performance for All Classifier Families}\label{FigureAllClassifierAll}
\hspace*{-1.5cm}
\centering
\includegraphics[scale=0.68]{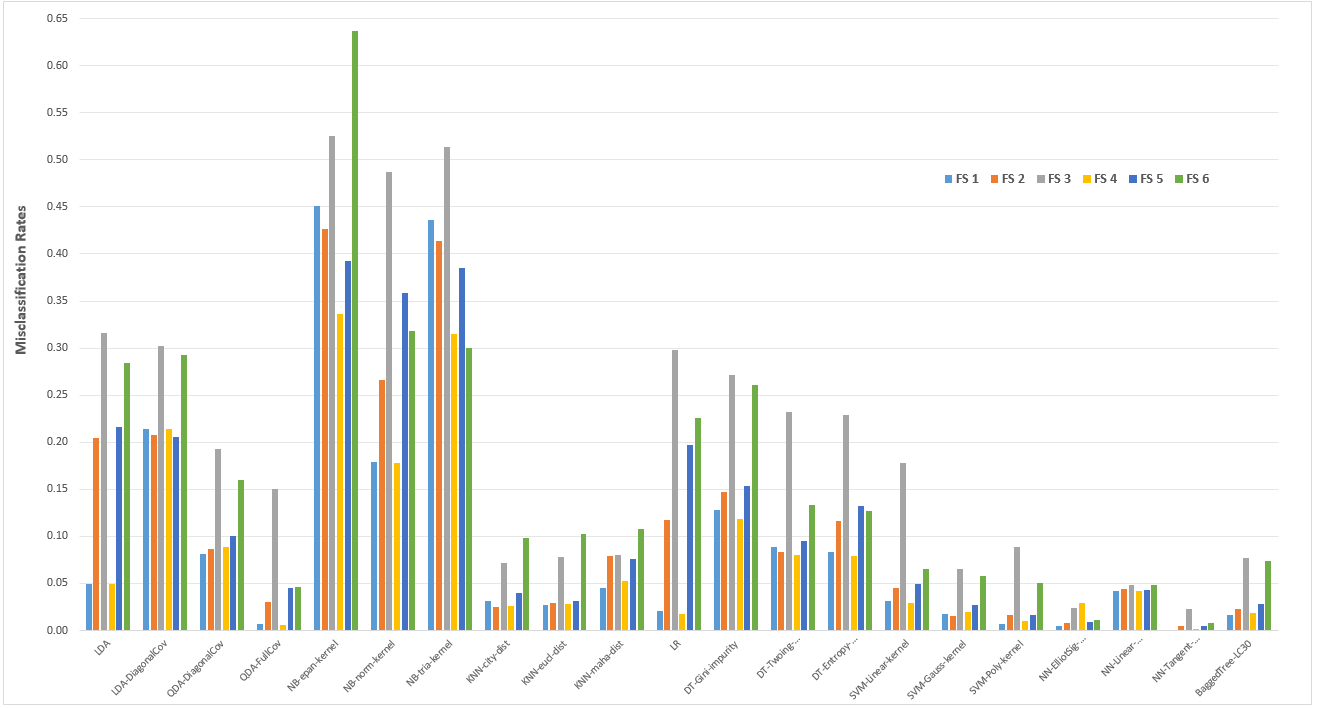} \\
\end{figure}

The main results of our paper are summarised in Figure  \ref{FigureAllClassifierAll}, where we have graphed the misclassification rates of each of the classifiers for each of the feature selections of Appendix \ref{sixmodels}, (indicated by a colour code), as well in Table \ref{FigureAllClassifierNumbers}, which lists the mean misclassification rate $\mu $ and its standard deviation $\sigma . $ %The latter are estimated by $K$-fold cross validation with $K=10 . $   
The parameters of the classifiers have been set empirically, so as to optimize the accuracy rates obtained after $K $-fold cross validation, while respecting the recommendations of the Machine Learning literature: see section \ref{section:Intra_Classifier_comparison} for further discussion of this point. Based on this figure and table, we can make the following observations.   
   
\begin{table}
\caption{Summary of Mean ($\mu$) and Standard Deviation ($\sigma$) Misclassification Rates estimated from $K$-fold Cross Validation}\label{FigureAllClassifierNumbers}
%\hspace*{-1.5cm}
\centering
\includegraphics[scale=0.9]{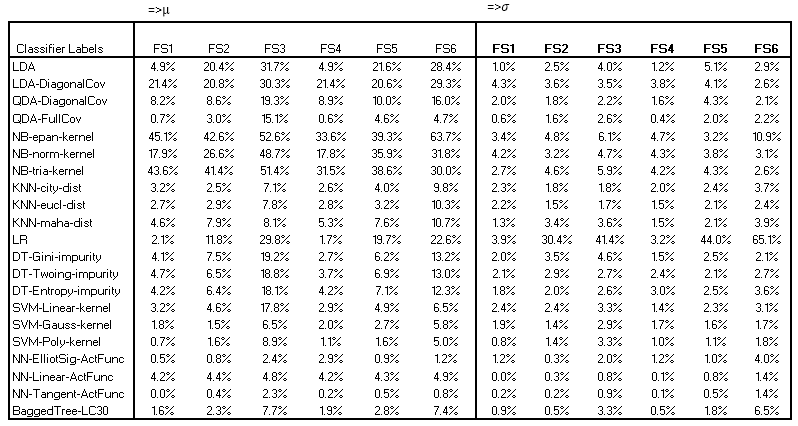} \\
\end{table}
\begin{enumerate}   
   
\item First of all, the figure indicates that the best performing classifiers are the Neural Networks, the Support Vector Machines and the Bagged Tree, followed by he $k $-Nearest Neighbour and the QDA classifiers, with Na\'I've Bayes overall the worst performer.     
   
\item To quantify this impression, we further aggregate the accuracy and misclassification rates for each Classifier, Following Delgado and Amorim (2014),  by computing the empirical mean and standard deviation of the accuracy rate over the different Feature Selections.   
These are recorded in Table \ref{RankingClassifiers}. According to this table, the best performing Classifier Families with the highest average accuray rates are, indeed, the Neural Network with the Tangent as Activation Function (mean accuracy 99.3\% and s.d. 0.6\%), the SVM with Polynomial Kernel (96.8\%, 1.6\%) and the Bagged Tree (96.0\%, 2.2\%). This is in line with results in the literature such as those of King {\it at al.} (1995) and Delgado and Amorim (2014). A bit surprisingly, perhaps, 'QDA-FullCov' also comes to the top league with a quite reasonable mean of 95.2\% and standard deviation of 1.6\%.   
     
\item If we focus on the effects of Feature Variable Selection, we see that for the majority of the classifiers families, the miss-classification rates for FS3 (in Grey) and FS6 (in Green), each with only two feature variables, are significantly higher than for the others. There are some exceptions, such as 'NN-Linear', where they all are comparable (and also relatively low) and 'QDA-FullCov', 'NB-norm-kernel', 'NB-tria-kernel', 'DT-Entropy', where FS6 does not perform too badly relative to the others.   
   
\item Still regarding Feature Variable Selection, across all classifiers, performances associated with FS1 (in Blue) and FS4 (in Orange) are very close in terms of misclassification rates, with the average across all classifiers (standard deviation in bracket) being respectively 8.5\% (2.0\%) and 7.5\% (2.0\%). A similar remark applies for the feature selections FS2 and FS5. Given that FS1 respectively FS2 require as additional feature the (most quoted) 5-year CDS rate, $s $, it will in practice be preferable to choose FS4 over FS1 or FS5 over FS2 because for a given nonobservable counterparty, chances are that it might not have any liquidly quoted CDS rates at all, including the 5-year rate. 
   
\item To justify our, literature-recommended, choice of $K = 10 $ (cf. Breiman {\it et al.} (1984) and Kohavi (1995)) for $K $-fold Cross Validation which we have used to compute the different empirical accuracy rates, we examined the dependence on $K $ of the empirical accuracy rates for the DA, LR and SVM families: see Figures \ref{tblDAKFold}, \ref{LRperformfig} and Table \ref{SVMmusigma}. We found that these rates do not vary much with $K $, which justifies assessing classifier performance with a pre-specified $K $, and that $K = 10 $ is a reasonable choice.   
      
\end{enumerate}

\begin{table}
\caption{Ranking of Classifiers based on Average Accuracy Rates and Standard Deviations across Six Feature Selections}\label{RankingClassifiers}
\hspace*{-0.5cm}
\centering
\includegraphics[scale=0.7]{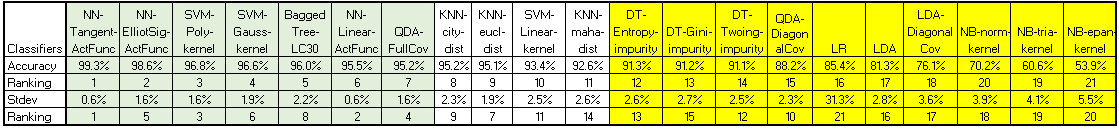}
\end{table} 

As mentioned earlier, it is common in a certain areas of Machine Learning, such as Image Recognition, to first perform some kind of Feature Extraction procedure such as a Principal Component Analysis (PCA), before passing to the classification stage. This is often done to reduce the dimension of the feature variable space. For us, the size of this dimension, being at most 16, is not much of an issue, but the presence of strong correlations between some or all of our individual feature variables might be. For classical regression this  has been well-investigated under the name of {\it Multicollinearity in Regression}. For financial and economic applications this is an important issue, as has traditionally been recognised in Econometrics: financial variables are typically known to be strongly correlated, particularly those having a term structure, such as the ones we use for our feature vectors: historical and implied volatilities, probabilities of default. Cross-sectionally, these may also be strongly correlated, such as for example Implied volatility of a certain maturity and the corresponding historical volatility. Figure \ref{PerfNBCorrelation} shows the empirical distribution of pairwise correlations across the 16 feature variables and indicates the presence of significant correlations in our data set.   
   
\begin{figure}
\caption{Summary of Classifier Performance and PCA ($K=10$ in Kfold, bandwidth $b^*=0.2$, Treesize $z^*=20$, $k^*=9$ for $k$NN, Learning Cycle $c=30$, all corresponding with Classifier Summary)}\label{ClassifiersPCAFigure}
\centering
\includegraphics[scale=0.68]{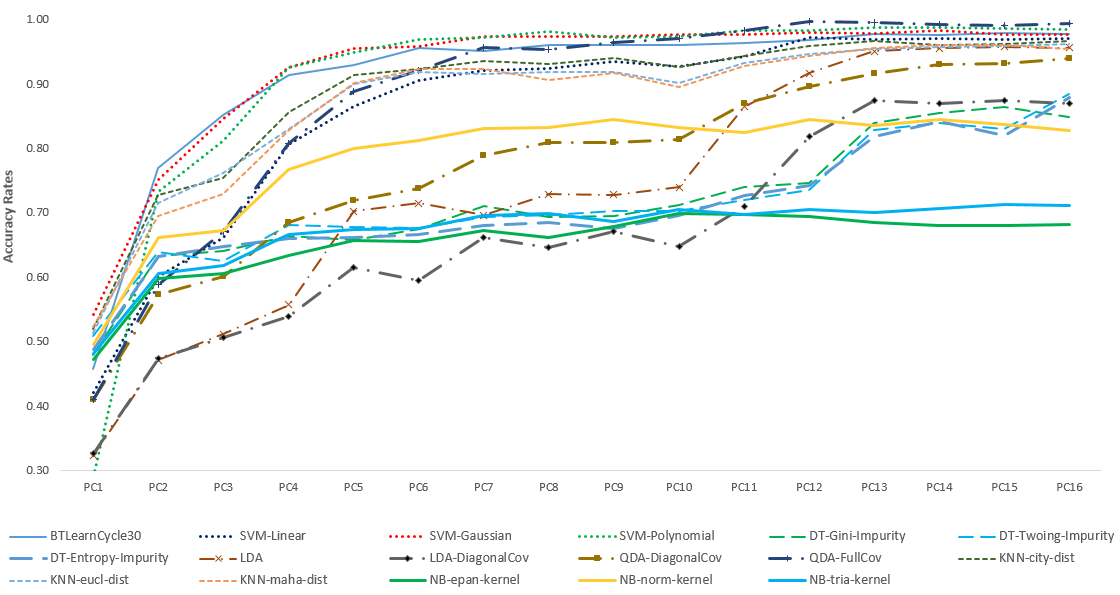}
\end{figure}

On the basis of this observation, we examined the impact of the correlations amongst our feature variables for six out of the eight Classifier families by replacing the original feature variables by those resulting from a preliminary Principal Component Analysis or PCA of our data set. Instead of the original feature variables of FS1 we have taken their coordinates with respect to the Principal Components as input for the classifications (which amounts to performing an orthogonal transformation in $\mathbb{R }^{16 } $), letting the number of PCs we use vary from 1 to 16. Finally, we performed a like-for-like comparison with classification based on FS1, the full vector of feature variables, by  comparing for each Classifier the classification performance calculated in PCA space with the one calculated with the original FS1. The idea is that the PCs are orthogonal, and thus uncorrelated, while the components of FS1 are not. If the classification results for both are similar, this shows, or at least is a strong indication, that multicollinearity is not an issue. Figure \ref{ClassifiersPCAFigure} and Table \ref{ClassifierPCAtbl} summarize our results:   
   
\begin{enumerate}
\item Figure \ref{ClassifiersPCAFigure} shows, as expected, that as more Principal Components are used for the Classification, the Accuracy Rates of the Classifiers increases, going to a maximum when all 16 components are used. It also shows that, with the exception of classifiers from the DA families, this performance roughly "flattens out" between PC5 and PC7. By contrast, some members of the two DA families require a greater number of PCs to come near their maximum accuracy rates, in particular QDA-DiagonalCov or LDA-DiagonalCOV. The situation of QDA-FullCov is more similar to that of the non-DA classifiers, in line with Table \ref{RankingClassifiers}. The rather brutal assumption of a diagonal covariance matrix in the QDA-DiagonalCov and LDA-DiagonalCOV  algorithms will of course already disregard any correlation information present in the data set, which may explain this anomalous behaviour.   

\item It is interesting to note that the first Principal Component (PC) already explains 98\% of the variance, and the first two 100\%, within two-decimal accuracy. Nevertheless, at least 5 PCs are necessary for the accuracy rates of the classifiers to stabilise, while the additional components only contribute infinitessimally to the variance. ''Variance explained'' is not a good predictor of classification accuracy, and one should be careful with using PC to reduce feature vector size.

\item Rather, PC should be used as a diagnostic test for signalling correlation effects. We did a like-for-like comparison between straightforward classification with feature vector FS1 and classification using the 16 PCs of FS1, by computing the differences between the respective Empirical Accuracy Rates (as always obtained by 10-fold cross validation). The results can be found in the final column of Table \ref{ClassifierPCAtbl} headed '$A(PC)-A(FS1)$'. We see that 
with the exception of NB and of LDA and QDA with diagonal covariance matrix for which, by assumption, correlation in the data is disregarded.   
Such correlation {\it would} be taken into account by a PCA, leading to a different classification. (Incidentally, the assumption of a diagonal covariance matrix becomes innocuous if feature variables are uncorrelated, such as the ones resulting from a PCA.)   
\end{enumerate}

\begin{table}
\caption{Classifier Performance, \% Variance Explained by PCs ($K=10 $ in Kfold, bandwidth $b^*=0.2$, Treesize $z^*=20$, $k^*=9$ for $k$NN, Learning Cycle $c=30$)} \label{ClassifierPCAtbl}
\centering
\includegraphics[scale=0.8]{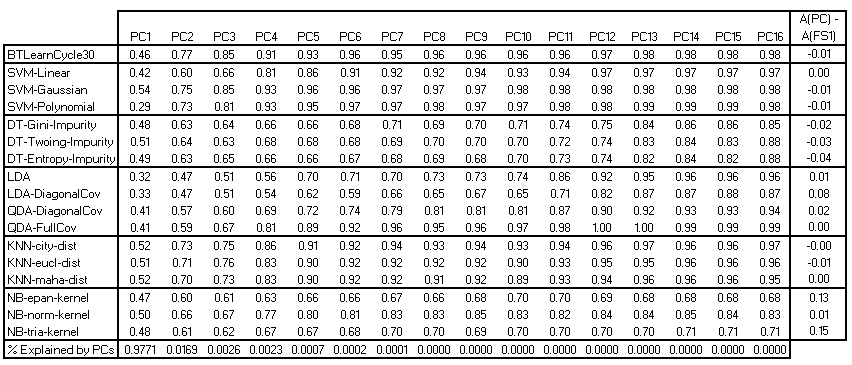}
\end{table} 

\subsection{Intra-classifier Performance Comparison Results}\label{section:Intra_Classifier_comparison}   
We next give a brief summary of our Intra-classifier performance results, with further details on individual classifier performances given by the graphs and tables of Appendix \ref{sectionindividaulclass}.   
   
\begin{enumerate}   
\item Within each Classifier family, there is a significant amount of performance variation across the different parameterisation choices as well as across the different feature selections.   
   
\item Regarding the \textbf{Discriminant Analysis} (DA) families, Figure \ref{DAAccuracy} compares the performances for the two types of DA classifiers, Linear and Quadratic, with the two different so-called {\it regularization choices} (Hastie {\it et al.}, 2009) for the covariance matrix: full versus diagonal,  for each of the six different feature variable selections (or {\it learning contexts} as they are also called).   

Figure \ref{DAAccuracy} shows the accuracy rates for QDA-FullCov and for LDA to be significantly larger than those of QDA-DiagonalCov and LDA-DiagonalCov, across all feature selections, while the standard deviations of their test errors are are either much smaller or approximately the same as for their Diagonal counterparts: cf. Table \ref{tblDA}.  Thus, using the full covariance matrix achieves a much better accuracy without introducing overfitting.   
   
\item Figure \ref{PerformanceNB} presents the Mean or Expected Accuracy Rates for \textbf{Na\"ive Bayes} classifiers as function of the band-width $b $, for the different kernel functions and the different feature selections. Na\"ive Bayes with $b \geq 0.3$ and feature selection FS6 and with either the normal or the Epanetchikov kernel underperforms all other classifiers studied in the paper, which partly motivated our Feature-Extraction study in subsection \ref{PCAstudyNB}.   
Table \ref{nbtbl2} lists the mean and standard deviation of the test errors as function of the bandwidth $b $ for the 18 Na\"ive Bayes classifiers we studied.   
Since there is no closed-form solution for the optimal choice of bandwidth $b^* $, we determine the latter empirically, based on performance estimates obtained from $K$-fold cross validation. Figure \ref{PerformanceNB} illustrates how: the average accuracy rate (over all classifiers) is found to be maximal for $b=0.2 $, and the graph shows that  the performance of ''norm6'' starts to  "fall off a cliff'' at $b=0.2 $, while performance of other classifiers either flatten out or also start to decline. Finally, for $b=0.1$, the performances for more than half of the classifiers are worse than for $b=0.2 . $   
 
\item Figure \ref{knnperffigure} and Table \ref{KNNperftbl2} show the accuracy rates for \textbf{$k$NN} for different choices of $k $ and different distance metrics, again for each of the six feature selections. There is again no analytical solution for choosing the optimal $k$ for $k$NN. Figure \ref{knnperffigure} indicates that the smaller the $k $, the better the accuracy rates we can achieve. However, small $k$ means that the set of nearest neighbours may be small, and that we will be taking the majority amongst a small number of counterparties, which may make the outcome sensitive to noise in the data. Jirina and Jirina (2008) and  Hassanat {\it et al.} (2014) recommend, as a rule of thumb,  that $k $ should be chosen near to $\sqrt{n } $ where $n $ is the number of training samples. In our study, $n = 100 $, corresponding to the 100 days leading up to Lehman's bankruptcy. Furthermore, we want $k $ to be odd to avoid ties in the Majority rule. As a result, we settled on $k^*=9$. As usual, Table \ref{KNNperftbl2} lists the mean $\mu$ and variance $\sigma$ of the test errors.   
      
\item Figure \ref{LRperformfig} and Table \ref{LRmusigmatbl} show the dependence of the empirical accuray rate on the number of strata in $K $-fold Stratified Cross Validation, for the six \textbf{Logistic Regression} classifiers, showing these to be quite stable. We note that accuracy for LR can be quite volatile, depending on the choice of feature variables, notwithstanding its popularity in the banking industry (credit scoring).   
   
\item Figure \ref{dtgraph1} and Table \ref{DTperformtbl} present the performances of the \textbf{Decision Tree} for the different  Impurity measures and different choices of Tree size (Maximum Number of Splits). If the resultant Decision Tree becomes too complex, it loses its interpretability and also tends to become unstable, according to Breiman {\it et al.} (1984). In our empirical study we settled on $z^*=20 $ as optimal Tree size since for larger $z $, performances become relatively flat while the complexity of the tree increases significantly.   
  
\item Figure \ref{SVMperformfig}   
graphs the performances of the \textbf{Support Vector Machine} classifiers for different choices of kernel functions, across the different feature selections. Tables \ref{SVMKFold} and Table \ref{SVMmusigma} respectively list the empirical Accuracy and Misclassification Rates and its standard deviation, as determined by $K $-fold Cross Validation and as a function of $K . $ They again justify our choice of $K = 10 $ as a reasonable one.   
      
\item Concerning the \textbf{Neural Network} (NN) classifiers, as for example reference \cite{Duda} emphasized, there is no foolproof method for choosing the number of Hidden Units of a Hidden Layer. We found empirically that, for our problem, the number of such units only had a limited influence on Intra-class performance variations: see Figure \ref{NNperformfig}. Note that the Accuracy Rates in this Figure are all rather high, between 93\% and next to 100\%. When reporting our Cross-classifier comparison results, we used a Hidden Layer Size of 10 units. Contrary to Layer Size, choice of Transfer Function has a bigger influence on performance, and we found in particular that the Tangent Sigmoid function achieved the best performance across all of the classifiers we investigated. Tables \ref{NNperftblAaccuracy} and \ref{NNperftbl} respectively list the numerical values of the Accuracy Rates of Figure \ref{NNperformfig} and the mean and standard deviation $\sigma$ of the test errors.   
   
\item Figure \ref{baggedtreeperffig} and Table \ref{baggedtreeperftbl} examine the performance variations of the \textbf{Bagged Tree} algorithm as function of the number $s $ of learning Cycles, starting with $s = 10 $, for each of the Feature Selections, as usual. They show that the empirical Accuracy and    Misclassification Rates vary little with $s $, and that a limited number of Cycles, between 15 and 20, is enough to achieve stable results. Bagged Tree was the third top-performer in our study; according to Hastie {\it et al.} (2009), the Bagged Tree algorithm, as an alternative to the Decision Tree, is more stable and performs better in out-of-sample tests. Our results in Figure \ref{BaggedTreegraph}, Figure \ref{baggedtreeperffig} and Table \ref{baggedtreeperftbl} confirm this.   
\medskip   
      
\end{enumerate}

\section{Conclusions} \label{SectionConclusion}
\subsection{Conclusions}
In this paper, we investigated CDS Proxy Construction methods using Machine Learning (ML) Techniques, based on publicly available financial market data, with the aim of addressing the Shortage of Liquidity problem for CDS rates. Machine Learning is already widely employed in the Pharmaceutical and Medical Sciences, Robotics, Oceanography, Image Recognition and numerous other domains. Ours is one of the first systematic study of ML applied to an important practical problem of Finance. From our results we can draw the following conclusions.

\begin{enumerate}   
\item After examining 156 classifiers from the eight currently most popular Classifier families in Machine Learning, we believe that carefully chosen classification algorithms with appropriate parameterization choices and feature variable selections can be used to construct reliable CDS Proxies with which to address the Shortage of Liquidity Problem for CDS rates described in the introduction. Such Proxy-constructions can achieve very high accuracy rates in cross-validation even if based on stressed financial data (in this paper, data from the 100 days leading up to the Lehman bankruptcy were used). Our top three performing classifier families  were the Neural Network, the Support Vector Machine and the Bagged Tree, a result which is consistent with  Machine Learning classification results using non-financial data reported in Kong {\it et al.} (1995) and Delgado and Amorim (2014).   
   
\item In contrast with existing studies such as Kong {\it et al.} (1995) and Delgado and Amorim (2014) which compared performances of classifiers on dozens and sometimes hundreds of non-financial datasets, we specialised our comparison to financial market datasets and for the purpose of one particular problem, that of CDS Proxy construction. This ensures that the performance comparisons are like-for-like. To the best of knowledge, ours is the first comprehensive classifier comparison based entirely on financial market data. Our findings for the overall ranking of best performing classifier families are nevertheless in line with those of existing literature in this area, with some exceptions, notably the Na\" ive Bayes classifiers. This can be explained by the particular characteristics of financial data, notably their in general highly correlated nature.   
   
\item We believe ours to be one of the first classification studies using only highly correlated data, and we have investigated the issue of Multicollinearity in  Regression (in the large sense) as it might affect the classification. Using Principal Component Analysis (PCA) as a Feature Extraction technique, we have shown that in our case, the correlations mostly do not impact strongly on classification; in particular, they do not for our three best performing classifier families. We recommend as good practice, when dealing with potentially strongly correlated features, to perform a PCA-space classification alongside classification using the untransformed or "raw" feature variables, both for validation and as a diagnostic tool.

\item We believe that ML-based CDS-Proxy methods to be superior to existing CDS Proxy Methods such as Curve Mapping or the Cross-sectional Regression: this methodology, by construction, meets all of the three criteria specified by Regulators, instead of only the first two of them (cf. the Introduction). Furthermore, and in contrast to the publicly available literature on Curve Mapping or Cross-sectional Regression, we have performed extensive out-of sample cross-validation tests for each of the ML algorithms we have investigated, thereby providing an objective basis for comparing these algorithms and selecting the best performing ones. Needless to say, this exercise should be repeated with other data sets to ensure reproducibility of our results, but their overall agreement with existing performance studies can be considered an encouraging sign.

\item A basic understanding of each of the ML algorithms is important, both to guide the parameter choices and for the interpretation of the empirical results. For this reason the paper has given an introductory presentation of each of the eight Classification families we have used, illustrated by a simple running example, thereby providing the necessary theoretical background in the specific context of a real-world problem.      
   
\item The paper has investigated the dependence on tunable parameters such as the number of nearest neighbours in $k$-Nearest Neighbour, the bandwidth in Na\"ive Bayes, the Tree Size parameter in the Decision Tree algorithms or the Hidden Layer size for Neural Networks. We also investigated the effect of varying the number of Strata in $K$-fold Stratified Cross Validation. In absence of theoretical results, the tuning of these parameters has to be done empirically, based on cross validation, while taking into account the recommendations of the existing literature.   
   
\item Our empirical study found that, despite its popularity within the Corporate Banking community, Logistic Regression is not amongst the top classifier families, and can be especially poor for classifications using only few features. Na\"ive Bayes also performed relatively poorly, in contrast with results for classification based on non-financial data. The reason for this should be sought in the Class Conditional Independence Assumption which underlies Na\"ive Bayes, and which is contradicted by the strong correlations which usually are present in financial data. A similar remark applies to the versions of the DA algorithms which restrict themselves to using diagonal covariance matrices only, and which can be considered as subclasses of Na\"ive Bayes.   
      
\end{enumerate}

\subsection{Future Directions}
Counterparty Credit Risk and Machine Learning are both active research areas, with the former driven by a dynamic business and regulatory environment and the latter by exciting technological progress. We would like to provide two potential directions for future research.   

First, our study is based on counterparty and feature variable data from the 100-day period leading up to Lehman's bankruptcy, and involving American investment-grade corporates from the Financial sector only: cf. Appendix \ref{sixmodels}. Our motivation was to assess the validity of the proposed CDS-proxy method in a "stressed'' economic environment with therefore, in principle, ''noisy'' data. As this paper shows, the proposed method works well in at least one example of such an environment, but the exercise should be repeated for other periods, both to ensure reproducibility of our conclusions and also because,  in practice, financial institutions are required to cover all sectors and all regions in both ''stressed'' and ''calm'' economic environments. It might in particular be interesting to investigate our CDS Proxy construction in an "exuberant" period, where investors may be less concerned with Counterparty Credit Risk, and market values of our feature variables might paradoxically become less reliable as predictors for credit quality.     
   
Furthermore, as already mentioned previously, the techniques discussed in this paper should also be useful in other areas of Finance. For example, classification techniques can be used to construct equity proxies for private firms that are not traded in the stock market by associating such firms with publicly traded firms on the basis of suitably chosen sets of feature variables. Such proxies can find useful applications in private equity investment and its risk management.

\appendix
\section{Features and Data}\label{sixmodels}

In this section, we present the six different sets of feature vectors, or Feature Selections, we used, and which we referred to as ''FS1-FS6'' in Table \ref{classifiersfig}. Following some of the Machine Learning literature, one also refers to such Feature Selections as Models. Please note that our study is not meant to be prescriptive in terms of feature selection. We have used our own empirical experience, as well as a literature survey regarding which financial variables are statistically significant, to predict CDS rates to arrive at the six feature vectors listed below. An alternative would have been to use Automatic Feature Selection, and we did experiment with Stepwise and Forward/Backward selection. However, rather than let the machine decide, we find it more sensible to decide ourselves which financial variables to use as features: we then can deliberately experiment with one selection versus another (taking more or less information onto account, using more or less maturities if the variable has a term structure, etc.). Moreover, automatic selection cannot know which features will be liquid or not: the features chosen by automatic selection will depend on the set of observed data which we use for training and cross-validation and which, by its very nature, will consist of observable counterparties only. For such counterparties typically more liquidly quoted financial contracts of various types will be available, and the selected features might be illiquid for the (set of) non-observable(s) one wishes to apply the trained algorithm to. This would then create a Shortage of Liquidity problem within the Shortage of Liquidity problem that we are trying to solve in the first place. This consideration incidentally also motivates experimenting with larger and smaller sets of feature variables, as we have done.   
\medskip   
   
\noindent {\bf Feature selections}:   
\smallskip   
   
\noindent \textit{FS1 } (see below for the meaning of the individual variables):   
\begin{equation} \label{c2sixmodeleq1}
\mathbf{x } = \left( s,PD_{6m},PD_{1y},PD_{2y},PD_{3y},PD_{4y},PD_{5y},\sigma_{3m}^{imp},\sigma_{6m}^{imp},\sigma_{12m}^{imp},\sigma_{18m}^{imp},\sigma_{1m}^h,\sigma_{2m}^h,\sigma_{3m}^h,\sigma_{4m}^h,\sigma_{6m}^h \right)   
\end{equation} \\
\textit{FS2:}
\begin{equation}
\mathbf{x } = \left( s,PD_{5y},\sigma_{6m}^{imp},\sigma_{4m}^h \right)   
\end{equation} \\
\textit{FS3:}
\begin{equation}
\mathbf{x } = \left( s,PD_{5y} \right)   
\end{equation} \\
\textit{FS4:}
\begin{equation}
\mathbf{x } = \left( PD_{6m},PD_{1y},PD_{2y},PD_{3y},PD_{4y},PD_{5y},\sigma_{3m}^{imp},\sigma_{6m}^{imp},\sigma_{12m}^{imp},\sigma_{18m}^{imp},\sigma_{1m}^h,\sigma_{2m}^h,\sigma_{3m}^h,\sigma_{4m}^h,\sigma_{6m}^h \right)   
\end{equation} \\
\textit{FS5:}
\begin{equation}
\mathbf{x } = \left( PD_{5y},\sigma_{6m}^{imp},\sigma_{4m}^h \right)   
\end{equation} \\
\textit{FS6:}
\begin{equation}\label{c2sixmodeleq6}
\mathbf{x } = \left( PD_{1y},PD_{5y} \right) .   
\end{equation}
Here we note that   
\begin{itemize}%[leftmargin=-.5in]   
   
\item In FS1-FS3 above, $s $ stands for the 5-year CDS rate; only the 5-year rate is included because this is typically the most liquid term of CDS trades. Wherever $s$ appears in the feature list, this CDS Rate is required in order the classification of a nonobservable. In absence of such a rate for a given non-observable, and if one nevertheless wants to use FS1 - FS3 for the classification, one can use a so-called 2-stage model, in which one first runs a regression of the 5-year rate $s $ against FS4, FS5 or FS6 on the class of observable counter-parties, and then uses this regression to predict $s $ for the non-observable. This $s $ can than be added to the feature list used to classify the non-observable using FS1, FS2 or FS6. (Since we would only be regressing the CDS rate of a single maturity we would not need to worry about introducing model arbitrages across the CDS term structure.) However, as noted in the main text, adding $s $ does not necessarily improve, and can occasionally even worse classification performance.        
   
\item $PD_{6m}$,$PD_{1y}$,$PD_{2y}$,$PD_{3y}$,$PD_{4y}$,$PD_{5y}$ denote the counterparty's probabilities of default over, respectively, a 6-month, 1-year, 2-year, 3-year, 4-year and 5-year time-horizon.   
   
\item $\sigma_{3m}^{imp}$,$\sigma_{6m}^{imp}$,$\sigma_{12m}^{imp}$,$\sigma_{18m}^{imp}$ are the counterpart's at-the-money implied volatility as computed from European call options on its equity with, respectively,   a 3-month, 6-month, 12-month and 18-month maturity. According to 
Berndt {\it et al.} (2005), %\cite{Duffie},   
implied volatilities have statistically significant explanatory power for CDS rates.   
   
\item $\sigma_{1m}^h$,$\sigma_{2m}^h$,$\sigma_{3m}^h$,$\sigma_{4m}^h$,$\sigma_{6m}^h$ denote the historical volatilities estimated from 1-year historical equity price returns for terms of 1-month, 2-month, 3-month, 4-month and 6-month respectively.
   
\item All the data used in our study are from Bloomberg\texttrademark, which is readily available to financial institutions. In the case where PD data are not available, one can consider using EDF\texttrademark as an alternative.   
   
\end{itemize}

For the empirical part of our paper, we focused ourselves on counterparties in the US financial industry. We constructed our data sample based on the observations taken during the 100 calendar days leading to the bankruptcy of Lehman Brothers for all constituents of CDX-NA-IG coming from the financial sector. 
\newpage

\section{Empirical Results for Individual Classifiers}\label{sectionindividaulclass}

\begin{figure}
\centering
\caption{\textbf{Discriminant Analysis} - Accuracy Rates for 24 Classifiers estimated from 10-fold Cross Validation}\label{DAAccuracy}
\includegraphics[scale=0.8]{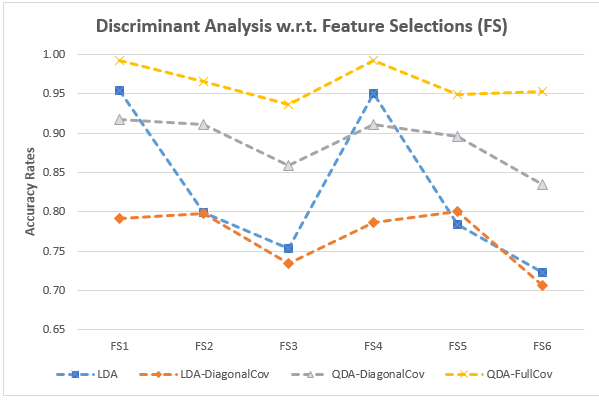} 
\end{figure}

\begin{table}
\caption{\textbf{Discriminant Analysis} - Means $\mu$ and Standard Deviation $\sigma$ of Test Errors for 24 Classifiers}\label{tblDA}
\centering   
\includegraphics[scale=1.0]{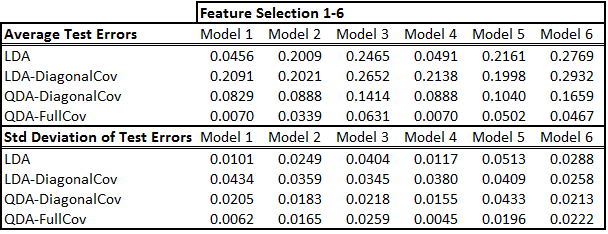} 
\end{table}
\begin{figure}
\caption{\textbf{Discriminant Analysis} - Performance w.r.t. to varying $K$ in $K$-Fold Cross Validation across Feature Selections}\label{tblDAKFold}
\centering   
\includegraphics[scale=0.8]{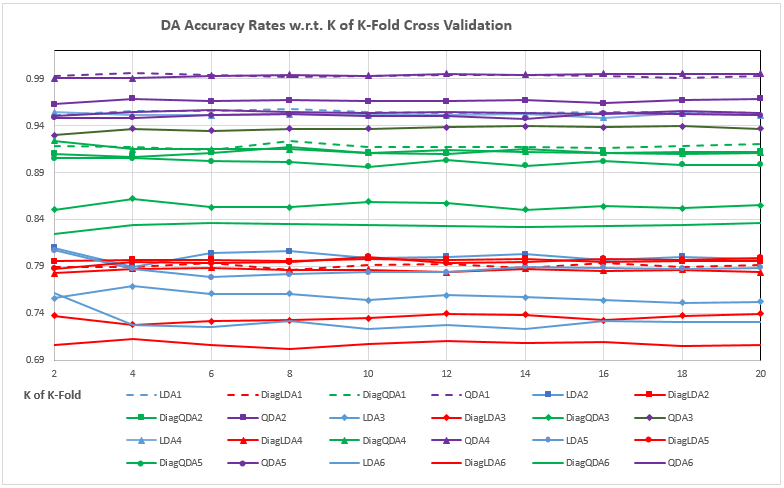} 
\end{figure}
\begin{figure}
\caption{\textbf{Na\"ive Bayes} - Expected Accuracy Rates w.r.t. bandwidth $b$ and Kernel functions estimated from $K$-fold Cross Validations (Optimal Choice of $b^*=0.2$)}\label{PerformanceNB}
\centering
\includegraphics[scale=0.9]{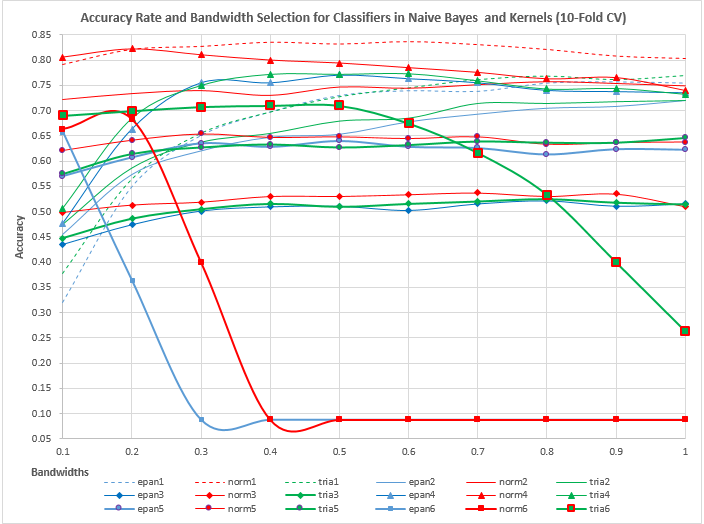} 
\end{figure}

\begin{table}
\caption{\textbf{Na\"ive Bayes} - Means $\mu$ and Standard Deviations $\sigma$ w.r.t. Bandwidths $b$ of Test Errors for 18 classifiers}\label{nbtbl2}
\centering
\includegraphics[scale=1.0]{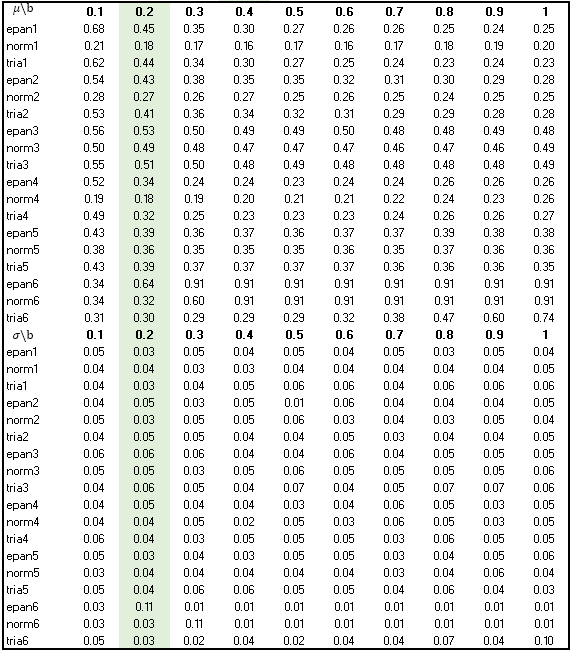} 
\end{table}
\begin{figure}
\caption{\textbf{Feature Extraction} - Highliy Correled Financial Feature Variables}\label{PerfNBCorrelation}
\centering
%\hspace*{-1.5cm}
\includegraphics[scale=0.6]{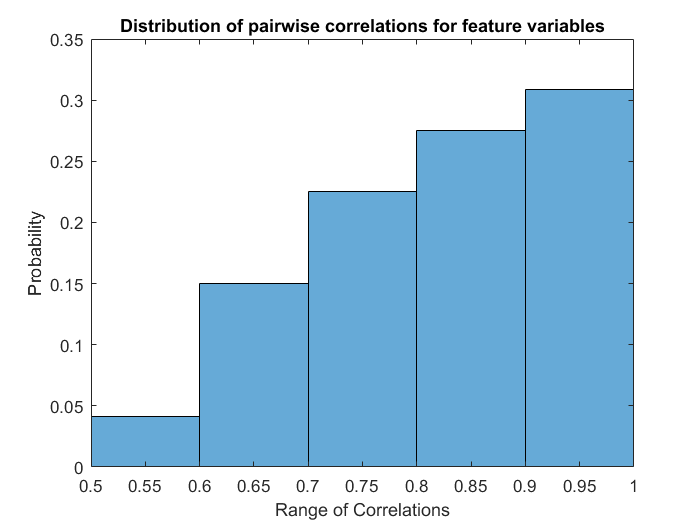} 
\end{figure}

\begin{figure}
\caption{\textbf{$k$NN} - Performance Variations for 18 Classifiers under  Classifier Family (Optimal $k^*=9$)}\label{knnperffigure}
\centering
\includegraphics[scale=0.7]{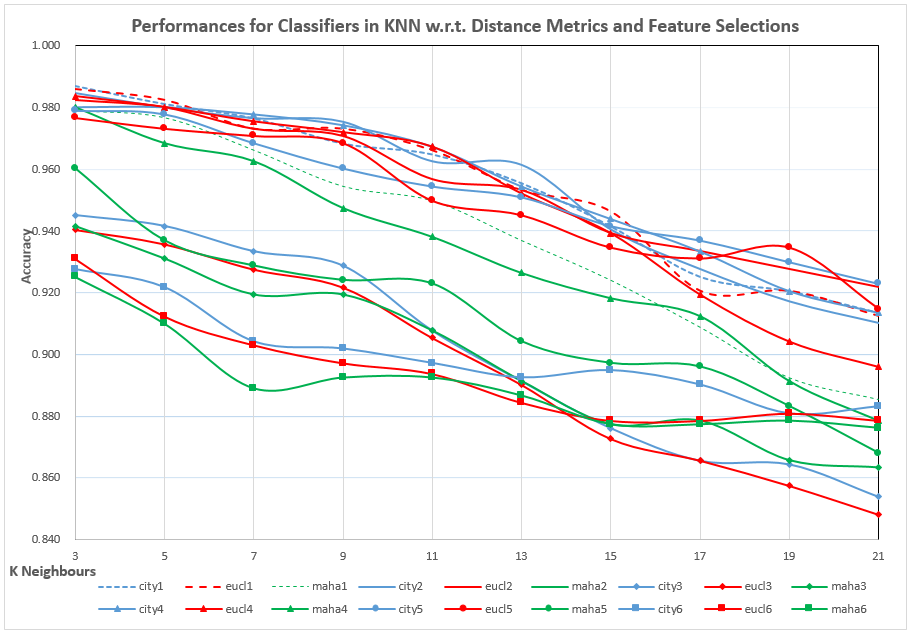} 
\end{figure}
\begin{table}   
\caption{\textbf{$k$NN} - Means $\mu$ and Standard Deviations $\sigma$ for Testing Errors for 18 $k $-classifiers; $k$ stands for the number of Neighbours}\label{KNNperftbl2}
\centering
\includegraphics[scale=1.0]{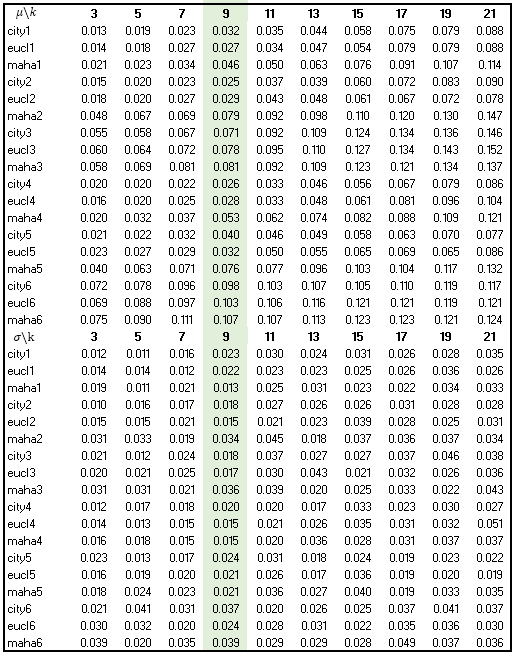}
\end{table}

\begin{figure}
\caption{\textbf{Logistic Regression}: Accuracy Rates w.r.t. varying $K$ in $K$-fold Cross Validation and across Feature Selections} \label{LRperformfig}
\centering
%\hspace{-0.6 cm}
\includegraphics[scale=0.8]{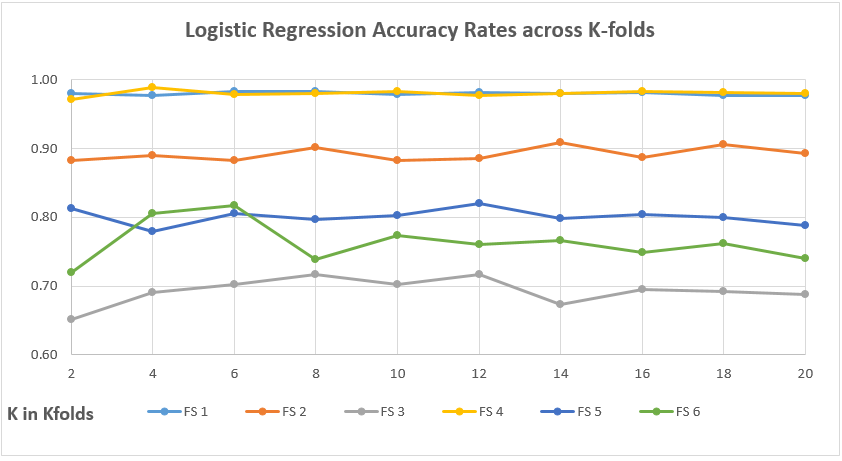}
\end{figure}

\begin{table}
\centering
\caption{\textbf{Logistic Regression} - Means $\mu$ and Standard Deviations $\sigma$ of Test Errors estimated by $K$-Fold Cross Validation}\label{LRmusigmatbl}
\centering
%\hspace{-0.6 cm}
\includegraphics[scale=0.8]{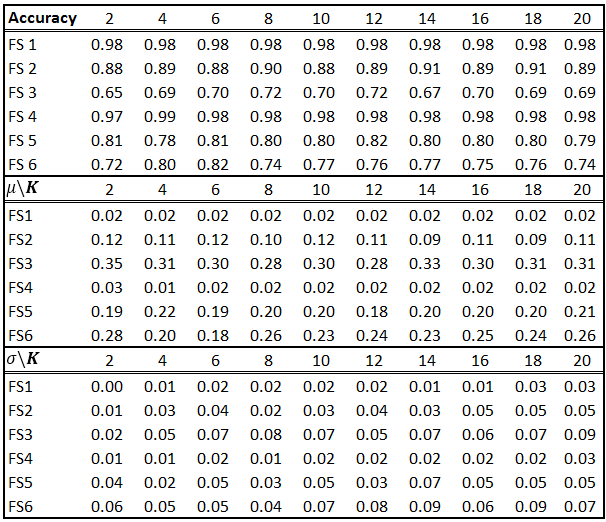}
\end{table}
 \begin{figure}
\caption{\textbf{Decision Tree} - Accuracy Rates w.r.t. Feature Selections, Tree Sizes $z$ and Impurity Measures (Optimal Tree Size $z^*=20$)}\label{dtgraph1}
\centering
%\hspace*{-1cm}  
\includegraphics[scale=0.8]{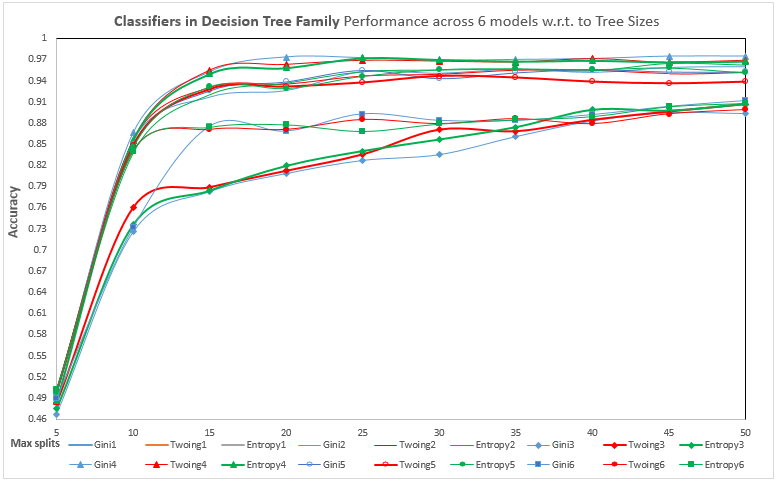}
\end{figure}

\begin{table}
\caption{\textbf{Decision Tree} - Means $\mu$ and Standard Deviations $\sigma$ of Test Errors w.r.t. different Impurity Measures, Maximum Splits and across Feature Selections (e.g., ''Gini1'' reads as the classification is conducted with Impurity Measure equal to ''Gini'' with FS1.)}\label{DTperformtbl}
\centering
\includegraphics[scale=0.85]{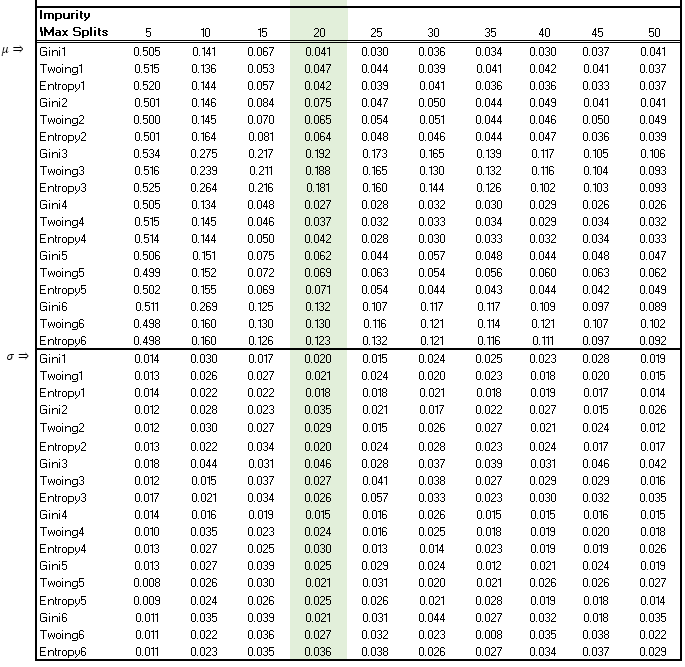}
\end{table}

\begin{figure}
\caption{\textbf{SVM} - Performance Statistics across Feature Selections}\label{SVMperformfig}
\centering
\includegraphics[scale=0.7]{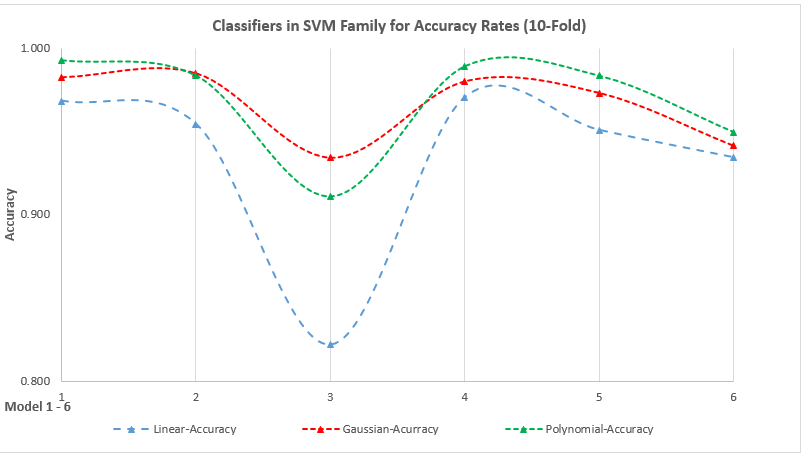}
\end{figure}
\begin{table}
\centering
\caption{\textbf{SVM} - Accuracy Rates w.r.t. varying $K$ of $K$-Fold Cross Validation}\label{SVMKFold}
\includegraphics[scale=0.8]{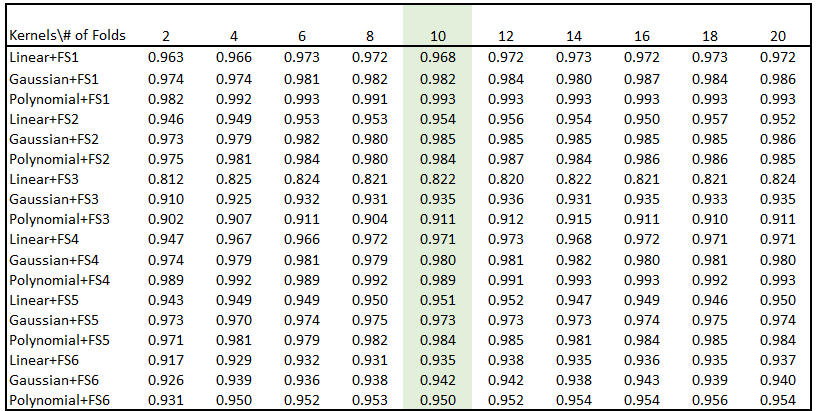}
\end{table}
\begin{table}
\centering
\caption{\textbf{SVM} - Means $\mu$ and Standard Deviations $\sigma$ of Test Errors w.r.t. to varying $K$ of $K$-Folds and across Feature Selections}\label{SVMmusigma}
\includegraphics[scale=0.8]{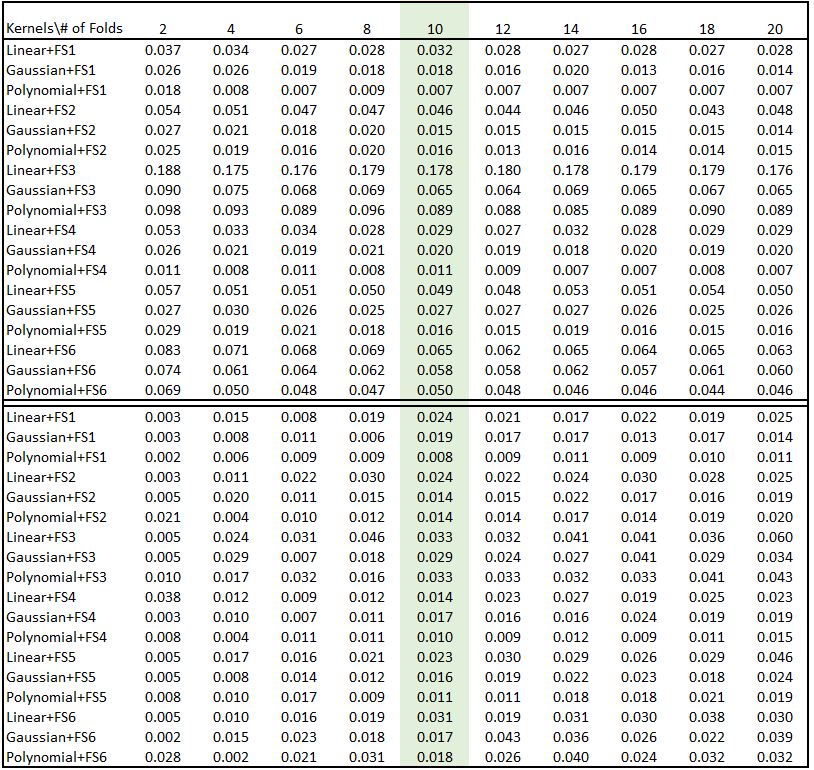}
\end{table}

\begin{figure}
\caption{\textbf{Neural Networks} - Response to Activation Functions, \# of Hidden Units, Feature Selections}\label{NNperformfig}
\centering
\includegraphics[scale=0.8]{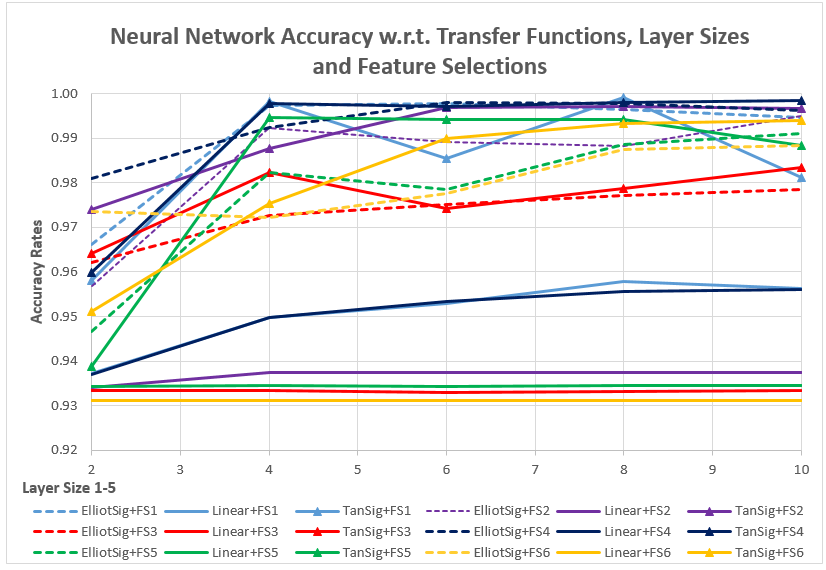}
\end{figure}

\begin{table}
\centering
\caption{\textbf{Neural Network}- Accuracy Rates across Activation Functions, Layer Sizes and Feature Selections (FS)}\label{NNperftblAaccuracy}   
\includegraphics[scale=0.8]{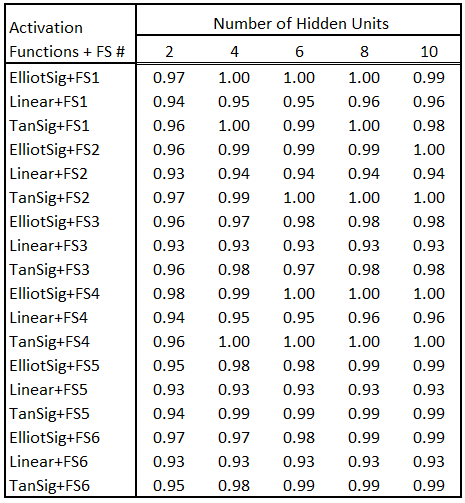}
\end{table}

\begin{table}
\centering
\caption{\textbf{Neural Network} - Means $\mu$ and Standard Deviations $\sigma$ of Test Errors estimated by $K$-fold Cross Validation where $K=10$ for different parameterization choices explained in Table \ref{NNperftblAaccuracy}}\label{NNperftbl}
\includegraphics[scale=0.8]{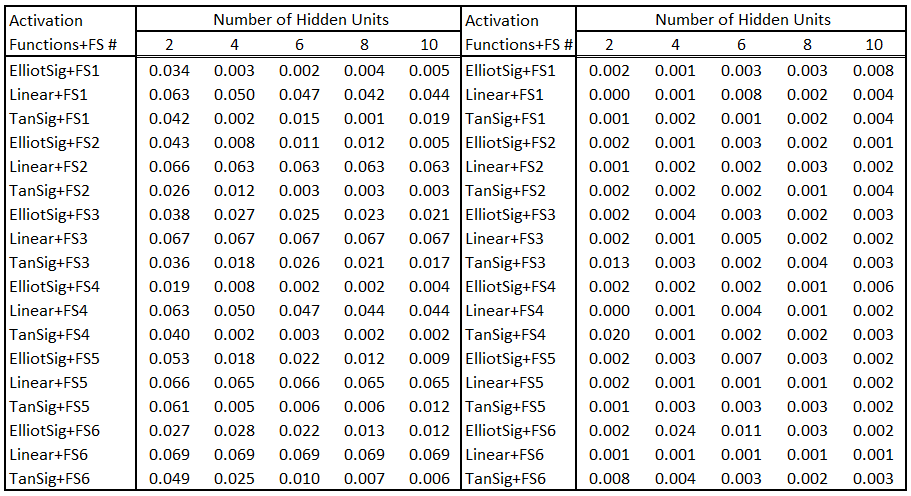}
\end{table}

\begin{figure}
\caption{\textbf{Bagged Tree} - Performances for w.r.t. Learning Cycles and Feature Selection}\label{baggedtreeperffig}
\centering
\includegraphics[scale=0.8]{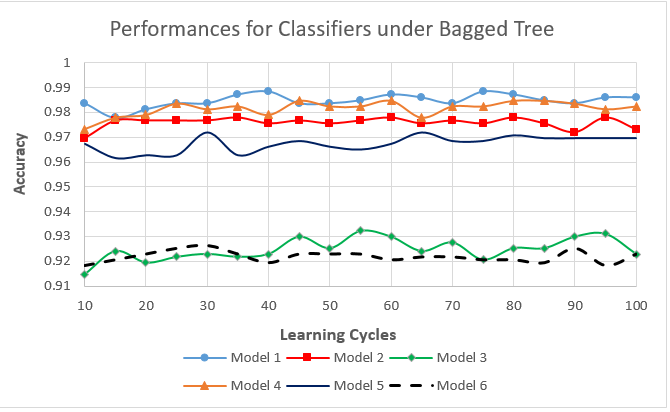}
\end{figure}   
   
\begin{table}   
\caption{\textbf{Bagged Tree}(An example for Ensemble) - Performance data w.r.t. Learning Cycles and Feature Selections (FS)}\label{baggedtreeperftbl}

\centering
\includegraphics[scale=0.7]{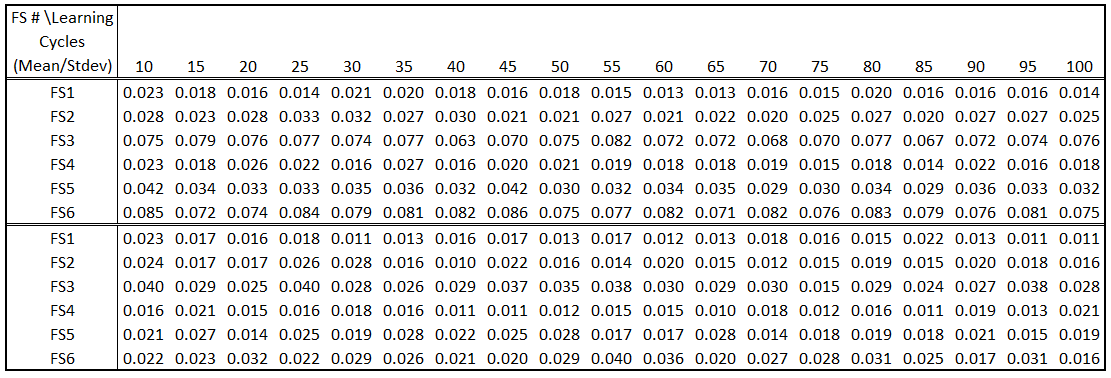}
\end{table}

\end{document}